\documentclass{aa}  

\usepackage{graphicx}
\usepackage{xcolor}
\usepackage{txfonts}
\usepackage[]{hyperref}

\begin{document}
\newcommand{\Euppk}{\mbox{$E_{\rm upp}/k$}}
\newcommand{\LSR}{\mbox{\em LSR}}
\newcommand{\vycma}{\object{VY\,CMa}}
\newcommand{\minky}{\object{M1--92}}
\newcommand{\tastar}{\mbox{$T_{\rm a}^*$}}
\newcommand{\tmb}{\mbox{$T_{\rm mb}$}}
\newcommand{\jk}{\mbox{$J_K$}}
\newcommand{\jn}{\mbox{$J_N$}}
\newcommand{\jkl}[2]{\mbox{$#1_{#2}$}}
\newcommand{\nuuu}{\mbox{$v_1$=1}}
\newcommand{\nudu}{\mbox{$v_2$=1}}
\newcommand{\nudd}{\mbox{$v_2$=2}}
\newcommand{\nutu}{\mbox{$v_3$=1}}
\newcommand{\nuz}{\mbox{$v$=0}}
\newcommand{\nhtzz}{\mbox{$0^-_0$}}
\newcommand{\nhtuz}{\mbox{$1^+_0$}}
\newcommand{\nhttz}{\mbox{$3^+_0$}}
\newcommand{\nhtdz}{\mbox{$2^-_0$}}
\newcommand{\nhttu}{\mbox{$3^-_1$}}
\newcommand{\nhtdu}{\mbox{$2^+_1$}}
\newcommand{\nhttd}{\mbox{$3^+_2$}}
\newcommand{\nhtdd}{\mbox{$2^-_2$}}
\newcommand{\jkk}{\mbox{$J_{K_{\rm a},K_{\rm c}}$}}
\newcommand{\jkul}[4]{\mbox{${#1}_{#2}$--${#3}_{#4}$}}
\newcommand{\jkkul}[6]{\mbox{$#1_{#2,#3}$--$#4_{#5,#6}$}}
\newcommand{\jkkl}[3]{\mbox{$#1_{#2,#3}$}}
\newcommand{\twopihalf}{\mbox{$^{2}\Pi_{1/2}$}}
\newcommand{\twopithalf}{\mbox{$^{2}\Pi_{3/2}$}}
\newcommand{\nutwo}{\mbox{$\nu_2$}}
\newcommand{\jtmdm}{\mbox{$J$=$3/2$--$1/2$}}
\newcommand{\rottenegg}{\mbox{OH\,231.8+4.2}}
\newcommand{\qxpup}{\mbox{QX\,Pup}}
\newcommand{\doceCO}{\mbox{$^{12}$CO}}
\newcommand{\sotwo}{\mbox{SO$_2$}}
\newcommand{\treceCO}{\mbox{$^{13}$CO}}
\newcommand{\treceC}{\mbox{$^{13}$C}}
\newcommand{\CdieciochoO}{\mbox{C$^{18}$O}}
\newcommand{\HtreceCN}{\mbox{H$^{13}$CN}}
\newcommand{\treceCN}{\mbox{$^{13}$CN}}
\newcommand{\CdoO}{C\mbox{$^{18}$O}}
\newcommand{\jdn}{\mbox{$J$=10$-$9}}
\newcommand{\jdsq}{\mbox{$J$=16$-$15}}
\newcommand{\jqc}{\mbox{$J$=15$-$14}}
\newcommand{\jsc}{\mbox{$J$=6$-$5}}
\newcommand{\jcc}{\mbox{$J$=5$-$4}}
\newcommand{\jno}{\mbox{$J$=9$-$8}}
\newcommand{\jtd}{\mbox{$J$=3$-$2}}
\newcommand{\jss}{\mbox{$J$=7$-$6}}
\newcommand{\jdu}{\mbox{$J$=2$-$1}}
\newcommand{\juc}{\mbox{$J$=1$-$0}}
\newcommand{\ammonia}{\mbox{NH$_3$}}
\newcommand{\vosio}{\mbox{$^{28}$SiO}}
\newcommand{\vnsio}{\mbox{$^{29}$SiO}}
\newcommand{\trsio}{\mbox{$^{30}$SiO}}
\newcommand{\sodos}{\mbox{SO$_2$}}
\newcommand{\tcsodos}{\mbox{$^{34}$SO$_2$}}
\newcommand{\protonwater}{\mbox{H$_3$O$^+$}}
\newcommand{\water}{\mbox{H$_2$O}}
\newcommand{\pwater}{\mbox{para-H$_2$O}}
\newcommand{\owater}{\mbox{ortho-H$_2$O}}
\newcommand{\waterdo}{\mbox{H$_2^{18}$O}}
\newcommand{\pwaterdo}{\mbox{para-H$_2^{18}$O}}
\newcommand{\owaterdo}{\mbox{ortho-H$_2^{18}$O}}
\newcommand{\waterds}{\mbox{H$_2^{17}$O}}
\newcommand{\owaterds}{\mbox{ortho-H$_2^{17}$O}}
\newcommand{\pwaterds}{\mbox{para-H$_2^{17}$O}}
\newcommand{\hydroxyl}{\mbox{OH}}
\newcommand{\rotten}{\mbox{H$_2$S}}
\newcommand{\gsim}{\raisebox{-.4ex}{$\stackrel{>}{\scriptstyle \sim}$}}
\newcommand{\lsim}{\raisebox{-.4ex}{$\stackrel{<}{\scriptstyle \sim}$}}
\newcommand{\psim}{\raisebox{-.4ex}{$\stackrel{\propto}{\scriptstyle \sim}$}}
\newcommand{\kms}{\mbox{km~s$^{-1}$}}
\newcommand{\s}{\mbox{$''$}}
\newcommand{\mloss}{\mbox{$\dot{M}$}}
\newcommand{\msun}{\mbox{$M_{\mbox \sun}$}}
\newcommand{\lsun}{\mbox{$L_{\mbox \sun}$}}
\newcommand{\my}{\mbox{$M_{\odot}$~a$^{-1}$}}
\newcommand{\ls}{\mbox{$L_{\odot}$}}
\newcommand{\ms}{\mbox{$M_{\odot}$}}
\newcommand{\mm}{\mbox{$\mu$m}}
\def\arcdeg{\hbox{$^\circ$}}
\newcommand{\seca}{\mbox{\rlap{.}$''$}}
\newcommand{\dega}{\mbox{\rlap{.}$^\circ$}}
\newcommand{\uv}{\mbox{\em uv}}
\newcommand{\aprop}{\raisebox{-.4ex}{$\stackrel{\propto}{\scriptstyle\sf
\sim}$}}
\newcommand{\apropg}{\raisebox{-.4ex}{$\stackrel{\Large \propto}{\sim}$}}
\newcommand{\about}{\mbox{$\sim$}}
\newcommand{\ttt}[1]{\mbox{10$^{#1}$}}
\newcommand{\hcop}{\mbox{HCO$^+$}}
\newcommand{\nnhp}{\mbox{N$_2$H$^+$}}
\newcommand{\CdiecisieteO}{\mbox{C$^{17}$O}}
\newcommand{\HtreceCOp}{\mbox{H$^{13}$CO$^+$}}
\newcommand{\HdoceCOp}{\mbox{H$^{12}$CO$^+$}}
\newcommand{\doceC}{$^{12}$C}
\newcommand{\SHAPE}{\mbox{\tt SHAPE}}
\newcommand{\shapemol}{\mbox{\tt shapemol}}
\newcommand\miguel[1]{{\color{red}MSG: #1}}
\newcommand\javier[1]{{\color{blue}JA: #1}}

   \title{M1-92: Asymptotic giant branch interruption and isotopic ratio paradox}

   \subtitle{Chemistry and morpho-kinematics from improved \shapemol\ modelling}

   \author{E. Masa
          \inst{1,2},
          J. Alcolea
          \inst{1},
          M. Santander-García
          \inst{1},
          V. Bujarrabal
          \inst{3},
          C. Sánchez Contreras
          \inst{4},
          A. Castro-Carrizo
          \inst{5},
          W. Steffen
          \inst{6},
          \and
          N. Koning
          \inst{7}}

   \institute{Observatorio Astronómico Nacional  (OAN-IGN), Alfonso XII 3, E-28014, Madrid, Spain
              \\
              \email{e.masa@oan.es}
              \and
              Facultad de Ciencias Físicas, Pl. de Ciencias 1, Universidad Complutense de Madrid (UCM), E-28040 Madrid, Spain
              \and
              Observatorio Astronómico Nacional (OAN-IGN), Apartado 112, E-28803, Alcalá de Henares, Spain
              \and
              Centro de Astrobiología (CAB), CSIC-INTA, ESAC, Camino Bajo del Castillo s/n, E-28691, Villanueva de la Cañada, Spain
         \and
             Institut de Radioastronomie Millimétrique, 300 rue de la Piscine, 38406 Saint-Martin-d’Hères, France
        \and
            Ilumbra, AstroPhysical MediaStudio, Kaiserslautern, Germany 
        \and
            Department of Physics and Astronomy, University of Calgary, 2500 University Drive NW, Calgary, AB T2N 1N4, Canada}
   \date{Received 30 May 2025 / Accepted 19 March 2026}

  \abstract
  {}   
   {The shaping of planetary nebulae on their evolution from asymptotic giant branch circumstellar envelopes to their final, most often axisymmetrical, form is still a process with many unknown details. The key to understanding the whole shaping process is the study of the transition objects called pre-planetary nebulae (pPNe). In this context, modelling tools must be kept to the standard of radio telescope capabilities, so we can make the most of the data they collect.}
    {In this work we first present the newest update of the \SHAPE\ and \shapemol\ modelling tools, adding ten new molecular species to be reproduced together with other general improvements. Later, we put this new update into practice to study \minky, a pPN with a rich chemistry that can provide valuable information on its origin and shaping.}
   {We created a 3D morpho-kinematical model of the nebula in \SHAPE\ that is able to reproduce 23 line profiles from the IRAM 30m telescope and HIFI/HSO and five maps from IRAM NOEMA. The observational dataset is reproduced simultaneously under the same physical conditions, adjusting only the relative abundance of the different species.}
   {We obtained a full description of the nebula's physical and chemical properties, and we provide the total estimates for mass (0.79 \msun), linear momentum (4.10$\times10^{39}$ g·cm·s$^{-1}$), and kinetic energy (6.48$\times10^{45}$ erg) as well as their detailed distribution across the nebula.  We also analysed the isotopic ratios, finding robust discrepancies (values of 10 versus 30) in the \doceC/\treceC\ ratio across structures depending on their age.}
{}

\titlerunning {New insights into M1--92 via \shapemol\ modelling}
\authorrunning {Masa, E.; Alcolea, J.; Santander-García, M.; et al.}
   \maketitle
   \nolinenumbers
%

\section{Introduction}
The transition from asymptotic giant branch (AGB) stars surrounded by their circumstellar envelopes (CSEs) to a white dwarf (WD) and their planetary nebulae (PNe) is a brief and spectacular metamorphosis experienced by low-to-intermediate mass ($\sim$0.8$-$8~\msun) stars, and it is still poorly understood, particularly regarding the physical mechanisms driving the transformation \citep[see][]{Balick_2002}. Far from simply being round fossils of AGB quasi-spherical stars and the isotropically expanding envelopes of AGB stars (AGB CSEs), $\sim$80\% of PNe depart from spherical symmetry and display a wide range of morphologies organised around (at least) one axis, whose origin we have been unable to grasp fully despite thorough efforts in the past decades (e.g. \citealp{Balick_2002}, \citealp{Jones_2017}, \citealp{Stanghellini_2016}). Similarly, the shaping of nebulae around intermediate post-AGB stars, not yet able to ionise the surrounding nebula, termed pre-planetary nebulae (pPNe), have proved to be challenging as well (e.g. \citealp{Sahai_1998}, \citealp{Ueta_2000}, \citealp{Sahai_2007}, \citealp{Lagadec_2011}, \citealp{Sahai_2011}, \citealp{Szczerba_2012}). It has become increasingly clear that a source of angular momentum provided from a binary or sub-stellar companion is a key ingredient to this puzzle, which almost certainly involves the launching of jets likely powered by accretion discs in rotation. However, the specifics of how this happens remain stubbornly elusive, with hydrodynamic/magnetohydrodynamic (HD/MHD) models struggling to match the vast zoo of PNe/pPNe shapes from mostly uncertain initial conditions (e.g. \citealp{Mastrodemos_1999}, \citealp{Icke_2003}, \citealp{Garcia-Segura_2005}, \citealp{Huarte-Espinosa_2012}, \citealp{Soker_2015}).

In this context, morpho-kinematic modelling of PNe/pPNe has proven to be a valuable tool for investigating their 3D spatial distribution and velocity field, which includes their orientation with respect to the plane of the sky and their kinematic (distance-dependent) age (e.g. \citealp{Solf_1985}, \citealp{Santander-Garcia_2004}, \citealp{Akashi_2013}, \citealp{Akashi_2018}). In this respect, the {\tt SHAPE} software \citep{shape} has become a standard in the PNe community, as it allows relatively easy modelling of these nebulae with custom geometries and a few parameters. Over the years, it has been used to successfully describe the present-day ejecta of a number of PNe displaying the most intricate morphology and kinematics (e.g. \citealp{Garcia-Diaz_2012}, \citealp{Santander-Garcia_2012}, \citealp{Clyne_2015}).

Following the release of \SHAPE, our group developed \shapemol\ (\citealp{Santander-Garcia_2015}), a plug-in that provides precise calculations of excitation and radiative transfer of rotational lines of \doceCO\ and \treceCO using the Large Velocity Gradient (LVG) approximation, as established by \cite{Castor_1970}. This tool generates synthetic spectral profiles and maps to match observational data. Simultaneously analysing low- and high-excitation lines allows for the determination of gas excitation conditions, more specifically, the density, temperature, and molecular abundance of each species relative to the total number of gas particles. From these data we can estimate the mass distribution in AGB envelopes and PNe, offering insights into molecular chemistry conditions and contributing to the refinement of formation models (e.g. \citealp{Santander-Garcia_2012}, \citealp{Santander-Garcia_2017}, \citealp{Doan_2017}, \citealp{Diaz-Luis_2019}).

As millimetre-range radio telescopes and interferometers improve in bandwidth and sensitivity, a plethora of spectral lines from species other than \doceCO\ and \treceCO\ are detected in observations of pPNe and PNe, allowing for a more complete description of the physico-chemical properties of the studied objects. Tracers such as HCN or \hcop, for instance, can reveal dense and hot pockets or shocked regions within the molecular envelopes of pPNe and PNe, while analysis of the spatial distribution of the HNC/HCN abundance ratio can be used as a diagnostic of ultraviolet radiation \citep{Viti_2002, Jorgensen_2004, Bublitz_2019, James_2020}. 

Bearing in mind these new possibilities to gain insight into pPNe and PNe formation, we have expanded {\shapemol} to include ten new molecular species and implemented a code to improve the direct matching of interferometric observations. Using these enhancements, we present new findings from the concurrent analysis of numerous IRAM-30m and HERSCHEL/HiFi spectra along with NOEMA interferometric maps focusing on different species within the pPN \minky.

The pPN \minky, also known as Minkowski's Footprint \citep{Herbig_1975}, is one of the most iconic objects of its class. First discovered by Minkowski ca. 1946 \citep{Minkowski_1946}, it consists of a bipolar reflection nebula around a B-type post-AGB central star, about 12\arcsec$\times$6\arcsec\ in size, with the major length following a conspicuous symmetry axis oriented at a PA $\sim$ 310\arcdeg\   \citep{Herbig_1975, Solf_1994, Bujarrabal_1998_shocks, Ueta_2007}. In the optical, the reflection nebula depicts a bi-lobed structure divided by a thick equatorial waist that partially obscures the south-east lobe. This orientation agrees with the expansion velocities measured in emission lines, which show approaching and receding movements in the north-west and south-east lobes, respectively \citep{Herbig_1975, Solf_1994}. These line emissions arise from a pair of compact knots placed along the symmetry axis, about 2\seca5--3\seca0 away from the centre, roughly in the middle of each lobe \citep{Herbig_1975, Solf_1994, Bujarrabal_1998_shocks}. They are detected in optical forbidden lines and vibrationally excited H$_2$ \citep[see][and references quoted before]{Davis_2005}, indicating the presence of active shocks.

The nebular composition is dominated by the molecular gas, which has been studied through observations of low-$J$ transitions of \doceCO\ and \treceCO, including interferometric maps at 0\seca5--1\seca5 spatial resolutions \citep{bujarrabal1997, Bujarrabal_1998_Dynamics, Alcolea_2007}. The molecular gas is located in the equatorial structure dividing the two lobes, in the walls of these two formations, and at the two polar ends of the nebula (the polar tips). In contrast to the optical images, because of the negligible dust opacity at millimetre-wavelengths, the emission from molecular gas also displays a remarkable equatorial mirror symmetry, i.e. both lobes are very similar to each other. A linear velocity gradient of 7.4\,\kms\ per arc-sec. was estimated, with expansion deprojected velocities up to 70\,\kms, which for a distance of 2.5\,kpc translates into a kinematic age of 1200\,a. This velocity gradient seems to hold in all directions, suggesting that the nebula was accelerated to its present velocities in a single brief (<100 a) event \citep{Alcolea_2007}. In addition to CO and H$_2$, the nebula has also been detected in other molecular species such as OH, SO, SO$_2$, and SiO, clearly indicative of O-rich composition, but also in CS, HCN, and \hcop, providing evidence of a rich and diverse chemistry \citep{seaquist1991, Alcolea_2022}. 

As stated before, in this paper, we fit the observational results of the four most abundant isotopologues of CO, \hcop, \HtreceCOp, HCN, and \HtreceCN\ from \minky\ while putting into practice the new capabilities of the upgraded \shapemol\ code. Details on these observations are given in Sect.\,2. In Sect.\,3, we describe the improvements in both {\tt SHAPE} and \shapemol\ developed for this work. Our new model for \minky\ is presented in Sect.\,4. Implications of this model on physical and chemical characteristics are discussed in Sect.\,5. Finally, the main results of this paper are summarised in Sect.\,6.

A very preliminary work was published in \cite{Masa_2024}, where a first approximation of the model and only some of the observational data were presented. Here we provide the final version of the model in addition to the complete work, from the new capabilities of \SHAPE \, and \shapemol \, to full analysis of the nebula's properties, and include all the observational data used.

\section{Observational data}

The data presented in this paper were obtained with the IRAM 30\,m-MRT, the IRAM NOEMA interferometer, and the HiFi instrument onboard the Herschel Space Observatory (30m, NOEMA, and HiFi hereafter). 

\subsection{Single-dish data}

The 30\,m data were acquired through various projects, namely 050--15, 160--15, and 047--16. They comprise the \juc\ and \jdu\ lines of the four most abundant CO isotopologues, \doceCO, \treceCO, \CdiecisieteO, and \CdieciochoO, and the \juc, \jdu, and \jtd\ lines of HCN, \HtreceCN, \hcop, and \HtreceCOp. Data were obtained with the Eight MIxer Receiver (EMIR) receiver in the 3\,mm, 2\,mm, and 1.3\,mm bands, using as a backend the fast Fourier transform units (FFTs) in its wide configuration, which provides a native spectral resolution of 200\,kHz. This results in a spectral resolution of 0.7 to 0.22\,\kms, although final spectra were resampled to a uniform resolution of 3.25\,\kms\ for increasing their S/N ratio, also facilitating the comparison with previous NOEMA observations and model results. The observations were performed in the wobbler switching mode (WSW), alternating between the ON and OFF positions at a frequency of 0.5\,Hz. The OFF positions are symmetrically taken 100\arcsec\ away in the azimuth axis to ensure a proper subtraction of the sky and instrumental contributions. This method provides very flat and stable baselines. Observations are automatically calibrated in \tastar\ units, which includes every 20\,mins estimates of the receiver noise and also corrections for the perturbations introduced by the atmosphere. Atmospheric absorption was corrected from the sky emissivity obtained in these calibration measurements and a model for the atmosphere at the 30\,m site, near Pico Veleta (Sierra Nevada, Spain), at an altitude of 2850\,m. Pointing was checked by the observation of quasars and other radio-continuum sources close in the sky to our target, about every two hours. After applying these corrections, we can assure that the pointing of the telescope is better than 2\arcsec. Relative calibration between the different runs were performed by the observation of strong spectral line emitters: specifically, the pPNe CRL\,2688 and the young PN NGC\,7027. For CRL\,2688 we adopted integrated area values of 218, 170, 56.5, 151, 324, and 300 K\,\kms (in \tastar\ scale) for \doceCO\ \juc, \doceCO\ and \treceCO\ \jdu\, and HCN \juc, \jdu, and \jtd, respectively. For NGC\,7027, we adopted integrated area values of 246, 238, 13.6, 9.8, 27.6, 21.5, 26.6, 49.6, and 42.0 K\,\kms (in \tastar\ scale) for \doceCO\ \juc, \doceCO\ and \treceCO\ \jdu\, HCN \juc, \jdu, and \jtd, and \hcop\ \juc, \jdu, and \jtd, respectively.
After this final recalibration, we adopt an absolute flux uncertainty of $\pm$5--10\% in the  3\,mm and 2\,mm bands, and of $\pm$10--15\% at 1.3\,mm. After eliminating bad and noisy scans, data from both horizontally and vertically polarised receivers were averaged. Zero- or first-degree polynomial baselines were also removed. Finally, all data have been rescaled into \tmb\ (K) using the main-beam efficiencies given in the EMIR user manual: 1.20 and 1.55 for CO \juc\ and \jdu\ respectively, and 1.15, 1.40, and 1.70 for \hcop\ and HCN \juc, \jdu, and \jtd\ respectively.

The HiFi/HSO data for the \jcc\ and \jno\ lines of \doceCO\ have been taken from the programme GT1\_dteyssie\_1 (PI: D. Teyssier), which have already been presented in the work of \cite{lorenzo2021}. Those for the \jss\ one have been taken from the project OT1\_vbujarra\_4 (PI: V. Bujarrabal). In this latter case, data calibration and reduction have been performed as in the case of the HIFIStars programme (PI: V. Bujarrabal), fully described in the work by \cite{alcolea2013}.

\subsection{Interferometric data}

NOEMA interferometric data, comprising the maps of the \jdu\ line of \treceCO, \CdiecisieteO, \CdieciochoO, HCN, \HtreceCN, \hcop, and \HtreceCOp, were obtained in project W17BJ. Data for CO isotopologues were acquired in the 1.3\,mm band (NOEMA band 3) while those for HCN, \hcop, and their \treceC\ substitutions were acquired using the 2\,mm receiver (NOEMA band 2). The observations were performed in the most extended A-configuration, together with the more compact C-configuration, resulting in final spatial resolutions of about 0\seca7. The PolyFix correlator was configured to provide an 8+8\,GHz full bandwidth coverage at a resolution of 200\,kHz. Final maps were also resampled to a spectral resolution of 3.25\,\kms. Observations happened with 9 antennas in March and April 2018 for two spectral setups, in A and C configurations: at the LO frequency 231.022 GHz (setup 1) five tracks were obtained, including one in C configuration, and at the LO frequency 168.022 GHz (setup 2) four tracks were observed, one in C configuration. In all the tracks a bright quasar was observed to perform the RF bandpass calibration. Two close calibrators (1901+319 and 1932+204) were observed every 25 min to determine instrumental phase and amplitude gain changes in time, and also to ensure a proper pointing and focus during the observations (every 25\,min or 50\,min). The main absolute flux calibrator in all the tracks was MWC349, with an adopted flux of 1.86 Jy at 219.6\,GHz and 1.64 Jy at 178.8 GHz (with S.I. 0.6). The flux calibration accuracy was found to be better than 10\% in both bands, considerably better at 2mm. Data calibration was made in CLIC by using the standard calibration procedures.

Whole band datasets were first mapped and then cleaned and inspected for line absorption/emission. After flagging channels affected by spectral signatures, data were channel-averaged in the {\em uv}-plane and imaged and cleaned once more for obtaining maps of the continuum emission at the different observed frequencies (156,204.7--164,310.9, 171,690.6--179,797.3, 219,188.8--227,303.8, and 234,930.9--242,795.8\,MHz). Pure (line-free) continuum {\em uv}-data were subtracted from the original datasets for obtaining pure (continuum subtracted) spectral line emission {\em uv}-data. Final datasets represented in LSRK-velocity (assumed systemic velocity of $-$0.5\,\kms\ in the LSRK frame) channels were obtained after correction for the corresponding rest frequencies (obtained from the JPL line catalogue) and resampling at a spectral resolution of 3.25\,\kms. This resample increases the signal-to-noise ratio and, given the velocity gradient in this source, translates into a spatial resolution along the line of sight direction of about 0\arcsec5, which is comparable to the spatial resolution of the interferometric maps.  In this paper, we show channel maps for a bandwidth of 140\,\kms\ that includes the whole detected emission. Image synthesis was performed using MAPPING within the GILDAS software\footnote{\href{https://www.iram.fr/IRAMFR/GILDAS/}{https://www.iram.fr/IRAMFR/GILDAS/}}. For the dirty mapping, we adopted a uv-plane sampling of 7.5\,m (half the diameter of the antennas), a pixel size of 0\seca1$\times$0\seca1, appropriate for the final attained spatial resolution, and a natural weighting scheme. Cleaning was performed using the Hogbom method, a gain of 0.15, and support polygons to restrict the area looking for clean components. We checked the convergence of the cleaning procedure by inspecting the cumulative cleaned flux. Finally, we also compared the total fluxes from the maps with those obtained with the 30\,m, concluding that there is no flux filtered out in the interferometric data as both measurements are compatible within the expected absolute calibration uncertainties. In this paper, we only present the \treceCO, \CdiecisieteO, \CdieciochoO, \hcop\ and HCN \jdu\ maps, with a resolution of around 0\seca50$\times$0\seca75, and a position angle (PA) of the beam of 76$^{\circ}$ for CO isotopologues and 86$^{\circ}$ for the other two species. These maps can be found in Figs. \ref{ratios1} to \ref{ratios5} in the Appendix, together with their moment zero maps and integrated emission spectra in \ref{mom0x5}\, and a tabulation of integrated fluxes from both IRAM 30m and NOEMA observations in \ref{tab:linefluxes}.

\section{Software}

\subsection{\shapemol\ update}
{\tt SHAPE} is a software code for the numerical simulation of the emission from gas nebulae \citep{shape}. The original version of {\tt SHAPE} is not intended for the modelling of the emission of molecular lines; this is done by the accompanying code {\shapemol} \citep[see][]{Santander-Garcia_2015}. In the version of {\shapemol} presented there, only the treatment of \doceCO\ and \treceCO\ was implemented. We introduce a new version of {\shapemol} to work along with the newest version of \SHAPE\footnote{\href{https://wsteffen75.wixsite.com/website/downloads}{https://wsteffen75.wixsite.com/website/downloads}}, which includes more molecular species and an updated treatment of the two most abundant CO isotopologues. The molecules chosen for this update are species relevant in the radio astronomy field but also with simple structure (linear molecules), where the vibrational modes can be ignored when calculating the radiative transfer at the studied frequencies (in the millimetre and sub-millimetre domains).

As stated previously, {\shapemol} is a complementary code for {\tt SHAPE} that computes synthetic line profiles and maps for the molecular line emission of a numerical nebula model. In summary, \shapemol\ solves the statistical equilibrium population of a given molecular species using the large velocity gradient (LVG) approximation formalism \citep[see][]{Castor_1970}, transforming the problem into a local one: the excitation properties of any molecular species at a given point does not depend on the properties of the rest of the nebula as a result of the different Doppler shifts. This calculation requires the simultaneous computation of several level populations that depend on both the collisional and radiative excitation transitions. Under the LVG formalism, since this is transformed into a local problem, the only relevant parameters are the density $n$, temperature $T$, relative abundance $X$ of the considered species, and the logarithmic velocity gradient $\epsilon = d \log(V)/ d \log(r) = (r/V)·(dV/dr)$, all referred to the particular point of the nebula for which the computations are being performed. The coupling of any given point with the rest of the nebula is controlled by the ratio of the macroscopic velocity to the local velocity dispersion: the larger the value, the more decoupled the different nebular points are. As a consequence, $\epsilon$ determines the effective escape probability of any given photon and so the importance of radiative trapping. LVG approximation applies to cases of an increasing radial velocity, which means an $\epsilon \geq 0$. This version of \shapemol\ has only been tested in these cases, and therefore, we recommend using it only for monotonously increasing radial velocities.

Although the LVG formalism was developed for large velocity gradients, i.e: large values of $\epsilon$, due to its simplicity, it is also applied for low $\epsilon$ values where, even though the expected accuracy in these cases is lower, its low computational cost and still very reasonable results, make it a very useful tool for a modelling process \citep[see][]{Bujarrabal_1994_a, Bujarrabal_1994_b, Bujarrabal_2013}.

\shapemol\ uses values of the absorption and emission coefficients, $k_v$ and $j_v$ respectively, of a set of rotational lines of a given molecular species. These values have been pre-computed for an ample range of densities, temperatures, relative abundances and values of $\epsilon$, and stored in a structured set of tables where {\shapemol} seeks the most appropriate values for a given nebular point (as a function of its local $n$, $T$, $X$, and $\epsilon$ values), and transfers them to the main code of {\tt SHAPE} for the production of synthetic spectral profiles and maps to be compared to real observations. Similar tables as the original ones are now available for \CdiecisieteO, \CdieciochoO, HCN, \hcop, HNC, SiO, CS, \nnhp, \HtreceCN, and \HtreceCOp. The tables for \doceCO\ and \treceCO\ introduced in the original work \citep[][]{Santander-Garcia_2015} have also been extended to match the new ranges of temperature, density and abundance covered by the new species (see Table \ref{tab:shapemol1}).

For the pre-computation of the molecular excitation (the $k_v$ and $j_v$ tables), we only considered collisional rates with H$_2$. Other collisional partners, such as He and electrons, as well as more complex molecular structures, such as the hyperfine structure, are not taken into account. Neglecting the hyperfine structure in \shapemol\ may have some impact in the \juc\ lines of HCN and \nnhp, resulting in slightly more opaque lines and narrower spectra. We note that in some sources, such as the one presented in this paper, this effect is not relevant due to the line width being dominated by the local turbulence and the source's kinematics. The collisional coefficient data are from the Lambda Database\footnote{\href{https://home.strw.leidenuniv.nl/$\sim$moldata/}{https://home.strw.leidenuniv.nl/$\sim$moldata/}} when they reached the temperature range we aimed for and from other more specific works otherwise. In the cases where separate tables for collisions with ortho-H$_2$ and para-H$_2$ are available, an ortho/para abundance ratio of three has been assumed. When not stated otherwise, Einstein coefficients and level multiplicities have also been taken from the Lambda Database, while the molecular constants used were provided by the Cologne Database for Molecular Spectroscopy (CDMS)\footnote{\href{https://cdms.astro.uni-koeln.de/}{https://cdms.astro.uni-koeln.de/}} and the National Institute of Standards and Technology (NIST)\footnote{\href{https://www.nist.gov/pml/molecular-spectroscopic-data}{https://www.nist.gov/pml/molecular-spectroscopic-data}} databases. (For further details on \shapemol's characteristics, see \cite{Santander-Garcia_2015}.)

For calculations of these pre-computed tables, the same non-local thermodynamic equilibrium (LTE) model based on the LVG approximation has been followed, and the general format has been maintained. Therefore, the absorption and emission coefficients have been calculated for ground-state rotational transitions from \juc \, up to $J$=17--16, for each combination of temperature, density and relative abundance, in the cases of $\epsilon$ being 0.2, 1 and 3. Temperature ranges from 5 to 1000\,K in steps of 5\,K and density from \ttt{2} to \ttt{12}\,cm$^{-3}$ in multiplicative factors of $\sqrt[4]{10}$ for all CO isotopologues, $\sqrt[8]{10}$ for SiO and CS, and $\sqrt[16]{10}$ for HCN, HNC, \HtreceCN, \hcop, \HtreceCOp, and \nnhp.

The relative abundance range has been chosen for each molecule so that the density column calculated for a radius of $10^{17}$cm covers those typically found in different objects, from AGB shells to planetary nebulae. The step chosen also has a multiplicative factor, this time of  $\sqrt[NT-1]{5.3\,\ttt{4}}$ for CO isotopologues, SiO, CS, \hcop, and \HtreceCOp, and  $\sqrt[NT-1]{\ttt{7}}$ for HCN, HNC, \HtreceCN, and \nnhp, where $NT$ is the number of different abundance values and so the number of tables provided. A total of 52 tables for each species are available, except for HCN, \HtreceCN, HNC, and \nnhp, whose typical range of values is about 2 to 3 orders of magnitude larger than for other species, and in which case 76 tables have been calculated, and SiO, calculated for a slightly higher abundance range with a total of 56 tables. This difference in number of tables aims to keep a similar abundance gap between tables. A summary of the characteristics and parameter coverage is shown in Table\,\ref{tab:shapemol1}.

Contrary to abundance, where different ranges require different iteration steps, the differences between multiplicative factors in density for each species are purely due to converging purposes. When it was needed, smaller density steps were taken to ensure smoother transitions between iterations and guarantee the full computation of each table. This is complemented by another strategy implemented in the temperature iterations, where the number of energy levels considered for calculation is reduced depending on the temperature considered, as lower temperatures will only significantly populate the lower levels, generating calculation noise from upper-level populations and increasing the computational cost.

The accuracy of these calculations has been checked for the most abundant isotopologues, assuming an analogue behaviour for their less common counterparts. These checks where done by comparing the opacities obtained during the computation of the $k_v$ and $j_v$ tables with those provided by RADEX\footnote{\href{https://var.sron.nl/radex/radex.php}{https://var.sron.nl/radex/radex.php}} \citep{Radex}, as well as through the result of radiative transfer expected in the simple analytic case of local thermodynamic equilibrium.

Differences with RADEX, which provides non-LTE molecular radiative transfer comparable to the LVG approximation in the case of a linear velocity gradient, are in the 5\% to 8\% range when the same collisional coefficients are used, which is the case of CO isotopologues, and between 5\% and 70\% otherwise. These discrepancies can be explained by the collisional coefficients data themselves, as differences of up to a factor 2 can be found between the different datasets used by RADEX and this work \citep[for analysis of both datasets for CS see][]{Denis-Alpizar_2018}. These comparisons were made only when collisional data were available for both RADEX and \shapemol. Temperature extrapolations or hyperfine structure were not considered when checking these results, as the huge discrepancies in those cases make the comparison irrelevant. Details on the differences between RADEX and \shapemol, as well as collisional coefficients data sources, can be found in Table\,\ref{tab:colls}.

\begin{table*}[]
        \caption{Summary of the results from the accuracy checks run on the pre-computed $k_v$ and $j_v$ tables together with the references of the collisional data used for each molecular species.}
    \begin{tabular}{|c|c|c|c|}
    \hline
    Molecular & RADEX: &   LVG test: & Collisional data\\ 
    species & max difference in $\tau$ (\%) &  max intensity error (\%) & source \\  \hline
    CO   & 8.3       & 6.10    & \cite{Yang_2010} \\ \hline
    SiO   & 45.3       & 5.96    & \cite{Balanca_2018} \\ \hline
    CS   & 69.7       & 3.41    & \cite{Schoier_2005}\\ \hline
    \hcop   & 6.5       & 0.98   & \cite{Schoier_2005} \\ \hline              
    HCN   & *       &  9.18     & \cite{Schoier_2005}\\ \hline
    HNC   & 27       & 10.50    & \cite{Schoier_2005}\\ \hline
    \nnhp   & *       & 0.99    & \cite{Schoier_2005}\\ \hline              
    \end{tabular}
    \tablefoot{The use of hyperfine structure by RADEX makes the comparison impossible for those species marked with *. The same source for the collisional data was used for all isotopologues of each molecule.}
    \label{tab:colls}
\end{table*}

\begin{figure}
\resizebox{\hsize}{!}{\includegraphics{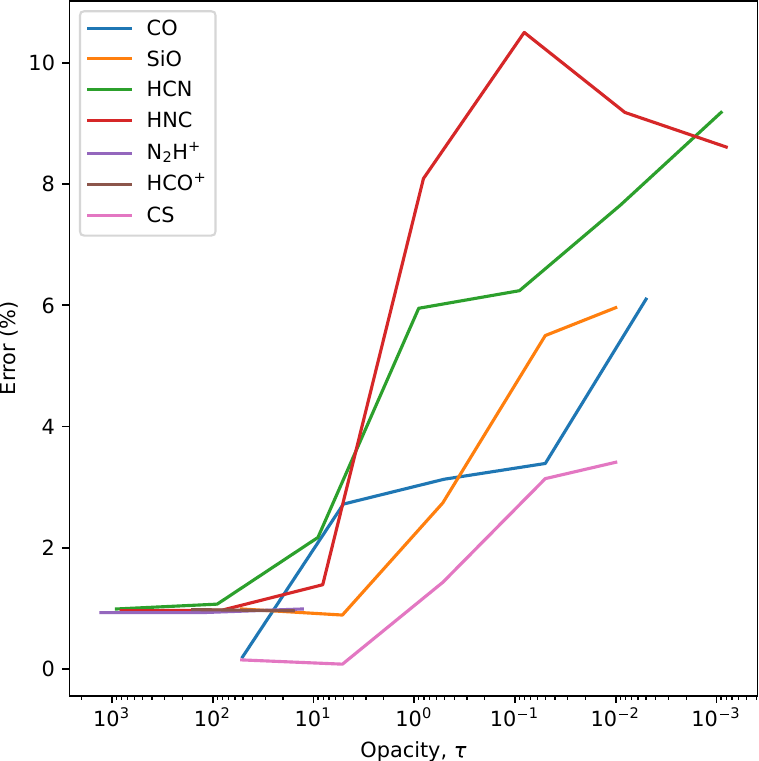}}
      \caption{Evolution of the error as opacity decreases, calculated as the discrepancy between \shapemol \, and the theoretical calculations with respect to the theoretical value. Only the opacity conditions where the molecule had thermalised were used.}
         \label{LVG_test}
\end{figure}

However, the most significant test of the final \SHAPE+\shapemol\ results is the comparison of its calculations in a case where we can analytically compute the solution. The test was run in a homogeneous thermalised sphere in which the kinetic temperature matches the excitation temperature of the molecule in those conditions while expanding with a linear velocity law. With Planck's law and the LVG approximation, the final intensity emitted at any point on the sphere can be obtained. In this case, the error is between 0.05\% and 10\%, generally decreasing as the opacity increases. This is the expected result, since as the opacity increases the lines tend to thermalise, converging to the kinetic temperature regardless of the approximation. Details on the errors for each molecule and opacity under the conditions required for the test are shown in Fig.\,\ref{LVG_test}. 

These new tables that expand and update \shapemol's capabilities can be downloaded via \SHAPE's website\footnote{\href{https://wsteffen75.wixsite.com/website/downloads}{https://wsteffen75.wixsite.com/website/downloads}} or the available electronic material of this work \footnote{\href{https://doi.org/10.5281/zenodo.19473007}{https://doi.org/10.5281/zenodo.19473007}}. For checking purposes, tables of \doceCO\ and \treceCO\ applied to LTE conditions are also available for the same ranges of temperature, density and relative abundance.

\subsection{SHAPE} 

Over the course of the expansion of {\shapemol}, the {\tt SHAPE} software has undergone a few upgrades in the frame of this work. Apart from fixing small errors and bugs, version BETA.2.0.0 has two new features that are useful for the simultaneous modelling of emissions from different transitions or molecular species. We briefly describe them below.

A new treatment of the cosmic microwave background (CMB) has been implemented in the Spectrum tab of the Render module. As in previous versions, by default, a black-body with a temperature of 2.73~K is assumed for the background, its emission being taken into account at each location in the first step of the radiative transfer calculation, and subtracted at the last step. In the new version, however, a background with a custom, user-defined temperature can be used instead, and the final subtraction can be disabled. This allows the modelling of sources against extended backgrounds with temperatures higher than the CMB.

{\tt SHAPE} now also includes a new scripting module that enables automation of the workflow. Employing its own simple scripting language, for which some examples are provided, the user can sequentially render a nebula in different transitions and save the resulting spectral profiles or map data cubes, for example, or explore the parameter space by automatically rendering a grid of models with different values of the desired parameters, and saving the results for later inspection. This new feature does not require {\shapemol} to work.

\subsection{Model-data comparison}
With every rendering, \SHAPE \, computes the emission along the line of sight through a ray-tracing algorithm. This produces a synthetic map with the same spatial resolution as the 3D model. This map then gets convolved with the beam provided for each transition and observing mode, creating the final dataset that is compared to the observational data. In the case of single-dish, the data is already ready for comparison, while for interferometric maps we added an extra step that subjects the raw data from \SHAPE \, to an `interferometrisation' process with the same conditions as the observation. For this process to be reliable, we must first SHAPE-model an area wide enough so there is no significant signal at the borders. This image should then be enlarged by a factor of two to avoid aliasing problems in the next step, and it is subsequently Fourier transformed. From these pseudo-visibilities, we only retained those matching the observational data {\em uv}-coverage, which were then processed with imaging synthesis techniques, similarly to the observed interferometric data.
This procedure ensures a direct comparison of model and real data in the image space even in the event of overesolution and lost flux problems (which are not the case here). This process is done through GILDAS's MAPPING software, after a conversion from \SHAPE's outcome file in TXT format into a FITS format file. Both scripts (`interferometrisation' and TXT to FITS conversion) are available via contacting the first author.

\section{M1--92 model}

Guided by the overall physical and chemical characteristics described in previous studies \citep{Solf_1994, Bujarrabal_1998_shocks, Bujarrabal_1998_Dynamics, Davis_2005, Alcolea_2007, Sanchez_Contreras_2008, Alcolea_2022}, we built a model of the pre-planetary nebula located 2.5~kpc away and oriented 40º with respect to the plane of the sky. Our de-projected model is 16\arcsec \, long with a cylindrical symmetry along this axis and a bipolar structure, with a radius of 2\seca25 in the equator and 0\seca5 at both ends of the nebula.  Thanks to \SHAPE's modelling features, an accurate position of the object, both in distance and orientation, as well as precise sizes, is easily reproduced.

\begin{figure*}
    \includegraphics[width=\textwidth]{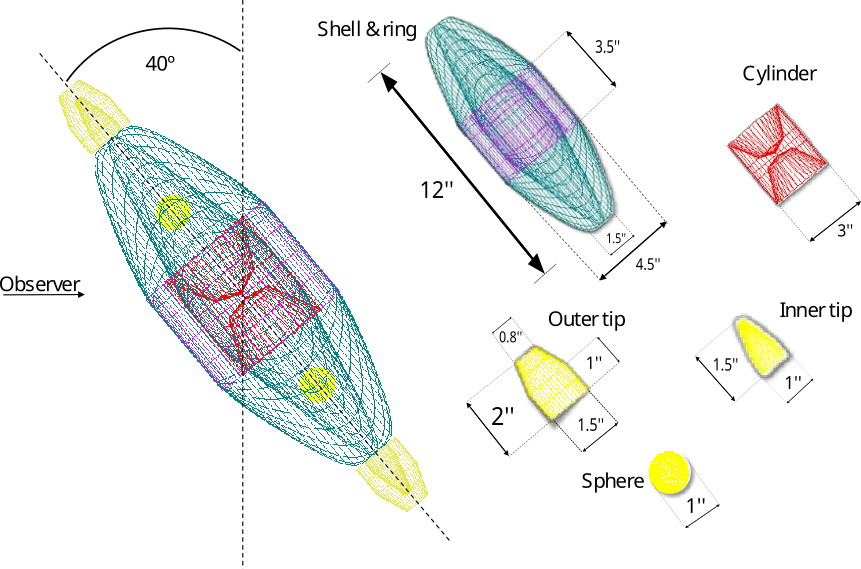}
      \caption{Our best-fit model in a wired representation showing the observer's point of view and its angle with respect to the plane of the sky. Each structure is shown in a different colour and separately: Green is for the outer shell, purple for the ring, red for the central cylinder, bright yellow for the spheres, and light yellow for both the outer and inner parts of the tips. The shell and ring are shown together, as the ring is only a limited region of the shell meant to define the locations of the species, but its physical properties are the same as in the shell. Values are provided for the main dimensions of each structure. Pictures of independent structures are not to the same scale. Note that 3.6$\times10^{16}$\,cm subtends 1\seca0 at the assumed distance of 2500\,pc.}
         \label{model}
 \end{figure*}

Our model (see Fig. \ref{model} for model building and Table \ref{tab:modelvar} for their physical properties) is made of six structures with revolution symmetry around the main axis and mirror symmetry with respect to the equatorial plane. They are divided into two categories: the outer ones, made of cold and dense gas with a shell-like shape and uniform physical properties, and the inner ones, made of warmer gas, with variable physical properties.

The largest component is the main shell, a prolate spheroid with flattened ends, which contains most of the inner (smaller) structures. This spheroid has been emptied by subtracting a similar smaller shape, leaving only a shell that will be filled with the equatorial structure, creating two empty lobes. Similar to all outer structures in our model, this shell is made of cold and dense gas, with constant values across all points, and low turbulence. Exact values of physical properties can be checked in Table \ref{tab:modelvar}.

A complementary structure to this is the ring. The ring is set in the equatorial area of the main shell, with 3\seca5 of length and the same thickness (0\seca75) and physical properties as the shell. Its only role in the model is to provide a defined region for those molecular species located only in the central part of the nebula when they reach the outer parts of its equator, but not extending along its shell.

Finally, the last outer structures are the outer parts of the tips. With very similar properties as the shell, these structures at each end of the nebula represent the continuation of that main spheroid when it squeezes into a structure 0\seca5 in radius.

The largest inner structure is the central cylinder. It is also 3\seca5 long and located in the nebula's centre, with a 1\seca5 radius and concave ends that complete the central shape of the empty lobes. This structure is associated with warmer but also denser gas than the outer shell, with both temperature and density laws ensuring the continuity of physical variables across the interface.

Finally, we have the two types of inner structures made of much warmer but much less dense gas. Those are the blobs and the inner part of the tips.
The blobs are represented by a sphere of 0\seca5 radius inside each lobe travelling along the main axis of the nebula. These are the only structures needing a much higher value for turbulence than the rest of the nebula to reproduce the velocity dispersion observed.
The inner part of the tips has similar physical properties to the blobs, but this time with turbulence more similar to the rest of the nebula, although still a bit larger. The full composition of the model and each structure's sizes are shown in Fig. \ref{model}.

For each of these structures, we set a turbulence, temperature, density, and velocity law, plus a different abundance for each molecular species present in them, in order to reproduce the observed line profiles and maps. All the structures, except the spheres, are given the same velocity law of a radial expansion proportional to the distance as \(V=\frac{64.75}{7.4}~\kms {\rm arcsec}^{-1}\times r({\rm arcsec})\), consistent with that detected in the nebula by previous studies. The spheres' velocity is defined by a step function in order to match the map observations, with the 0\seca7 closest to and the 0\seca3 farthest from  the centre of the nebula expanding radially (from the nebula's centre) at 55~\kms  and  at 15~\kms\, respectively. This is the equivalent to a front-like shock distribution, which stays consistent with the extreme velocity dispersion found across different species. The remaining physical variables set in the final model are shown in Table \ref{tab:modelvar}, and the final relative abundances of each species used in each structure are shown in Table \ref{tab:Abundances}. A more detailed description of the model and its structures can be found in the \SHAPE \, file of the model, provided among other additional material of this work.

\begin{table*}[]
        \caption{Physical variables for each structure in the best-fit model.}
    \begin{tabular}{|c|c|c|c|}
    \hline
    Structure & Temperature law $\pm$ Error (K) & Density law $\pm$ Error (cm$^{-3}$)  & Turbulence FWHM $\pm$ Error (\kms)  \\ \hline
    Shell \& Ring   & 17$\pm$2       & (7.0$\pm$1.9)\,\ttt{4}    & 2 \\  \hline
       &       & \(1.75\,\ttt{5}\) for \(r<0.5\)    &  \\
    Central cylinder & \(max(\frac{160}{6},\frac{160}{6r+1})\) & \((-0.7r+1.95)\ttt{5}\) for \(0.5<r<1\) & 2 \\
      & & \(7.0\,\ttt{4}\) for \(r>1\) & \\
      & $\pm$8$\%$ & $\pm$29$\%$ & \\ \hline
    Spheres   & 500$\pm$40       & (3.5$\pm$1.3)\,\ttt{4}   & 18$\pm$3  \\ \hline
    Tips in   & 600$\pm$48       & (7.5$\pm$2.4)\,\ttt{3}    & 5$\pm$1 \\ \hline
    Tips out   & 20$\pm$2       & (5.0$\pm$1.4)\,\ttt{4}   & 2 \\ \hline

    \end{tabular}
    \tablefoot{Turbulences under channel resolution (3.25\,\kms) are imposed. All errors presented take into account the instrumental errors and the main degeneracies of the model. See additional details in section 5.4.}
    \label{tab:modelvar}
\end{table*}

\begin{table*}[]
        \caption{Relative abundances with respect to collisional particles for each molecular species and model structure in the best-fit model.}
    \begin{tabular}{|l|c|c|c|c|c|c|}
    \hline
    Species & Shell &  Ring  & Central cylinder & Spheres & Tips in & Tips out\\  \hline
    \doceCO   & \multicolumn{6}{ |c| }{(5.5$\pm$1.1)\,\ttt{-4}} \\ \hline
    \treceCO   & \multicolumn{6}{ |c| }{(1.8$\pm$0.1)\,\ttt{-5}} \\  \hline
    \CdiecisieteO   & \multicolumn{6}{ |c| }{(1.0$\pm$0.1)\,\ttt{-6}}\\ \hline
    \CdieciochoO   & \multicolumn{6}{ |c| }{(6.0$\pm$0.2)\,\ttt{-7}}\\  \hline
    \hcop   & (5.0$\pm$0.3)\,\ttt{-9} & (2.5$\pm$0.1)\,\ttt{-8} & (1.0$\pm$0.1)\,\ttt{-8}      & (1.8$\pm$0.1)\,\ttt{-7}   & (1.6$\pm$0.1)\,\ttt{-6} & (5.0$\pm$0.3)\,\ttt{-9}\\ \hline
    \HtreceCOp   & (5.0$\pm$0.3)\,\ttt{-10} & (2.5$\pm$0.1)\,\ttt{-9} & (1.0$\pm$0.1)\,\ttt{-9}      & (1.8$\pm$0.1)\,\ttt{-8}   & (1.6$\pm$0.1)\,\ttt{-7} & (5.0$\pm$0.3)\,\ttt{-10}\\ \hline 
    HCN   &   \multicolumn{2}{ |c| }{<5.0\,\ttt{-9}}    &  (1.5$\pm$0.1)\,\,\ttt{-8}  & (3.5$\pm$0.2)\,\,\ttt{-7}   & (1.5$\pm$0.1)\,\,\ttt{-6} & <5.0\,\ttt{-9}\\ \hline
    \HtreceCN   &   \multicolumn{2}{ |c| }{<1.0\,\ttt{-9}}   &  <3.0\,\ttt{-9}  & <7.0\,\ttt{-8}   & <3.0\,\ttt{-7} & <1.0\,\ttt{-9}\\ \hline 
    \end{tabular}
    \tablefoot{Only the error of \doceCO\, abundance includes both degeneracy and instrumental errors. The rest of errors are presented relatively to the best-fit values of the model, taking into account only each individual abundance error margin due to observational data errors, since the degeneracy with density for \doceCO\, translates to all other values when affecting the physical variables of the model.} 
    \label{tab:Abundances}
\end{table*}

The 3D model was run inside a grid defined by a field of view of 40\arcsec\ $\times$ 40\arcsec\ and a resolution of 127 $\times$ 127 cells for line profiles and 255 $\times$ 255 cells for maps. Each rendering is done by filling the model with one species at a time and for a specific transition. In the case of single-dish line profiles, the results are convolved with a circular gaussian beam 3\% bigger than the nominal one of the telescope to account for tracking errors. For interferometric maps, we follow the procedure described in section 3.3, which ensures the same clean-beam geometry as in the observations. In every case, the render mode used was HD, with a CMB set at 2.73\,K and being subtracted. The spectral resolution used was 1000 spectral bands dividing a velocity range of $\pm70\,\kms$. We set the number of channels for interferometric maps to 35, through a range of $\pm55.25\,\kms$ (i.e. a velocity spacing of 3.25\,\kms). We also used the newly added scripting module to compute all the transitions sequentially, allowing for a myriad of combinations of different physical variables to be computed semi-automatically. This way we could explore a wide range of values for each parameter in a reasonable time, to finally find the best-fit model. The script used, together with the \SHAPE \, file of the model, can be found among the examples provided in the webpage.

We considered the model to be successfully built once it was able to reproduce all observations across all transitions and species using the same values for its physical properties. This means that when we add a new species to the model, the only adjustments allowed are the values of its relative abundance in each structure, with no changes in any other variable. All transitions, both from maps and line profiles, undergo a visual inspection by eye to ensure they are fit simultaneously. Previously, especially during model building, a mathematical inspection is done by calculating the emission ratio between model and observations, both for single-dish and interferometric maps. However, the final decision must be a compromise between each line's emission distribution, across all lines of the same molecular species for the relative abundance, and across all observational data for the common physical properties, making the human eye the most powerful tool to determine the fit. This method is widely used and accepted within the research community, particularly regarding morpho-kinematical modelling of complex nebulae \citep[e.g.][]{Solf_1985,Santander-Garcia_2004, Santander-Garcia_2008, Huckvale_2013}, and it has lead to solid results, such as the almost-perfect alignment between the orbital plane and the equatorial waist of every post-common-envelope planetary nebulae where both orbital parameters and nebular morphology have been studied, the latter relying on the human eye as fitting tool \citep[see Fig. 8 in][ and references therein]{Hillwig_2016}.

\subsection{CO model}

Among our observational data, CO is the molecule that can trace the overall structure of the nebula defined by the cold and dense gas, so it is be the main source of information to build our model. We have single-dish data of five \doceCO\ rotational transitions: \juc, \jdu, \jcc, \jss \, and \jno. For \treceCO, \CdiecisieteO \, and \CdieciochoO \, we also obtained line profiles of the first two transitions, \juc \, and  \jdu, as well as interferometric maps of the \jdu \, transition from each of them. From the interferometric data, particularly those obtained for the \treceCO \, transitions, with the brightest detected emission, we obtained an accurate description of the sizes and geometry of the outer structures of the nebula. The largest and densest structures are best studied from the low-J mapped transitions, and hence this is the case for the outer shell and the central area in the nebula. In contrast, we could best determine the physical conditions from the emission of the observed \doceCO\ transitions due to their larger variety of involved energies. Thus, we were able to detect emission from all structures, including the blobs, where, for instance, no CO emission has been detected in any low-J transition so far and hence for which the excitation conditions have not been properly studied until now.

From the modelling analysis, we deduced that the central cylinder has the most complex physical properties among the structures that compose it. Its temperature law is defined as an inverse function, decreasing as the radius increases until it reaches the temperature of the outer shell. A similar tendency is also set on the density, but this time with a linear decrease between a higher density centre and the outer shell. As a result, our model has a very small volume with both high density and temperature, which will make it relevant across all transitions covered by our observational data. This property distribution was chosen as a compromise to best fit the data in all observed transitions, even though it produces a peak too narrow in higher lines and excessive brightness in intermediate ones.

To achieve an accurate reproduction of the emission from all \doceCO\ transitions, we needed to introduce a double-layered structure of the tips. For the emission from these structures to be clearly detectable in the \jno \, line but absent in \jcc \, and lower, this double layer is a must. The hotter area detected in higher lines must be contained in a smaller and less dense part, while a denser but colder shell covers it. The latter ensures that while it has no relevant contribution to the lower line profiles, its presence is still clear on the maps.

The only structures for which not much information is provided by any of the CO observations are the spheres inside the lobes, introduced to reproduce the observed blobs. While their importance for the highest \doceCO \, line observed is clear, they are barely noticeable in the rest of the lines, contributing only by slightly broadening the emission further along in the velocity axis due to its high turbulence. In addition, most of their velocity range overlaps with that of the tips and even the central parts of the nebula, which results in an infinite number of possible combinations to obtain the final intensity observed in the line profiles. Lacking any further information, a proper assessment of its physical conditions as well as a final definite description of the model was not possible until the HCN and \hcop \, maps, both in \jdu, were evaluated.

 \begin{figure*}
    \includegraphics[width=\textwidth]{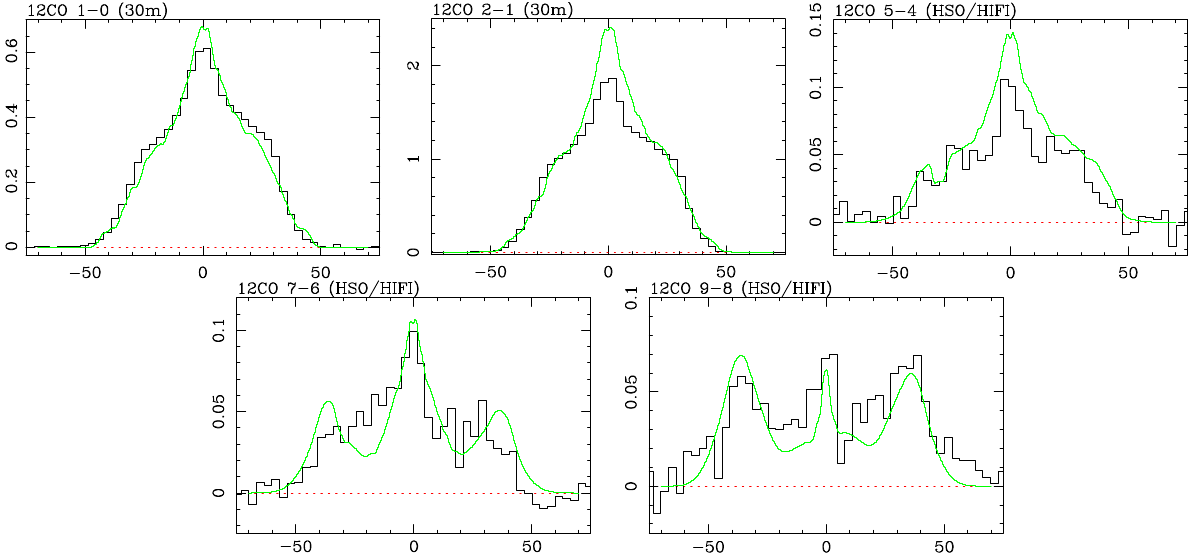}
      \caption{Comparison of line profiles obtained from single-dish IRAM-30m and HSO/HIFI observations (black) and model reproduction (green) on \doceCO \, lines in \tmb\ (K) vs LSR velocity (\kms).}
         \label{12COLines}
 \end{figure*}

 \begin{figure*}
    \includegraphics[width=\textwidth]{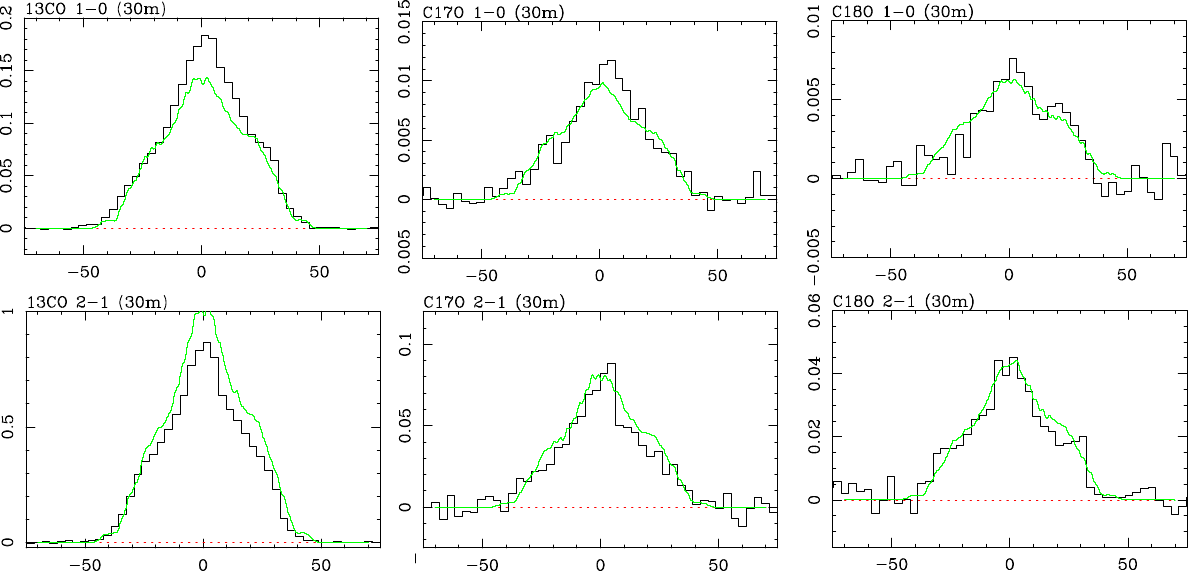}
      \caption{Comparison of line profiles obtained from single-dish IRAM-30m observations (black) and model reproduction (green) on \treceCO, \CdiecisieteO \,, and \CdieciochoO \, lines in \tmb\ (K) vs LSR velocity (\kms).}
         \label{COisolines}
 \end{figure*}

\begin{figure*}
    \includegraphics[width=\textwidth]{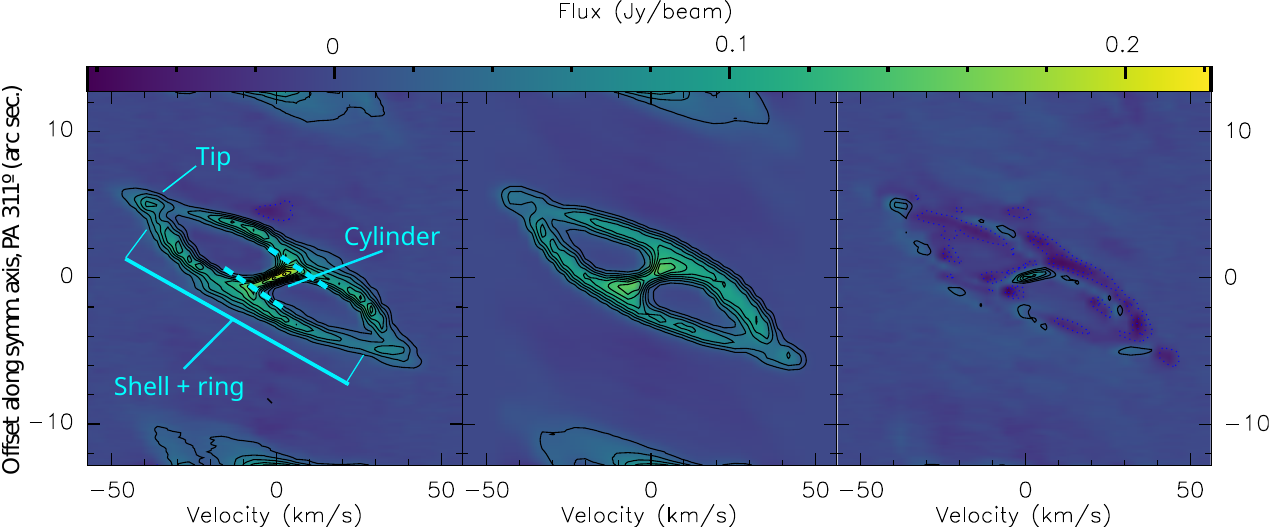}
      \caption{Position-velocity diagrams of observational data from IRAM-NOEMA (left), the model (centre), and residuals (observation - model; right) from the \treceCO \, \jdu \, map. Contours are drawn at 20\,mJy\,beam$^{-1}$ intervals. The structures that dominate the emission are annotated in the observational data panel.}
         \label{13comap}
 \end{figure*}

Figure \ref{12COLines} shows the comparison of \doceCO\ line profiles from observational data (in black) and model simulations (in green). We can see the same results for the line profiles of \treceCO, \CdiecisieteO \, and \CdieciochoO \, in Figure \ref{COisolines}. In all of these lines, the central peak is always the less well-fitted part of the model. Not only in intensity, the distribution of the peak differs slightly from the observations in all lines, especially in the highest-excitation ones, where the central peak of the observational data is displaced towards positive velocities. A look into the \treceCO \,position-velocity (P-V) diagram in Fig. \ref{13comap} shows us that the source of this inconsistency is found in the central area of the nebula. 

\begin{figure*}
    \includegraphics[width=\textwidth]{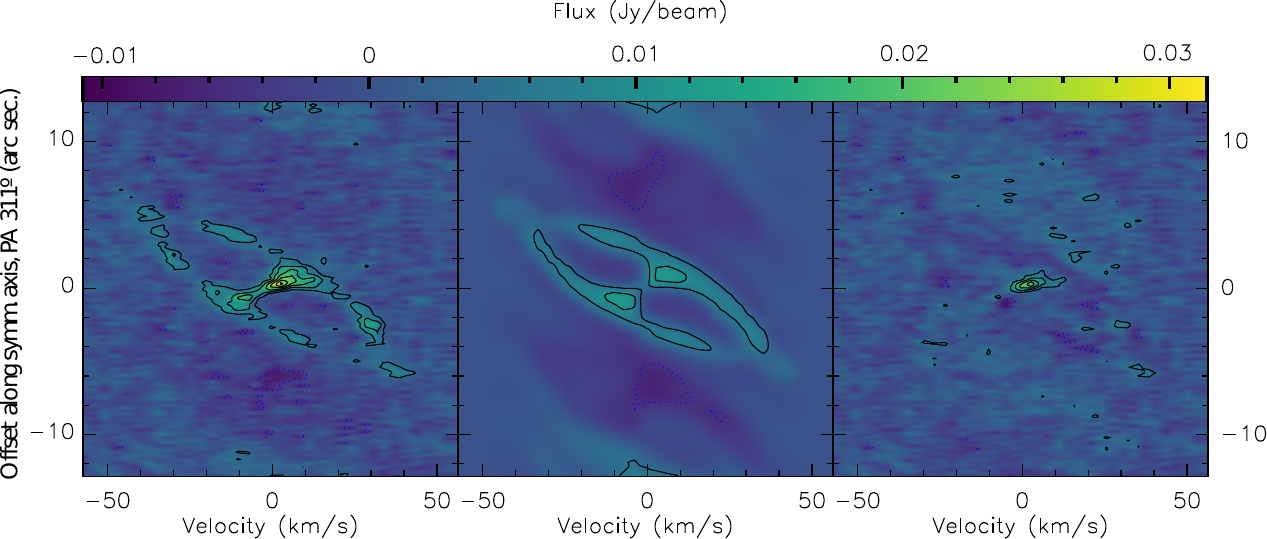}
    \includegraphics[width=\textwidth]{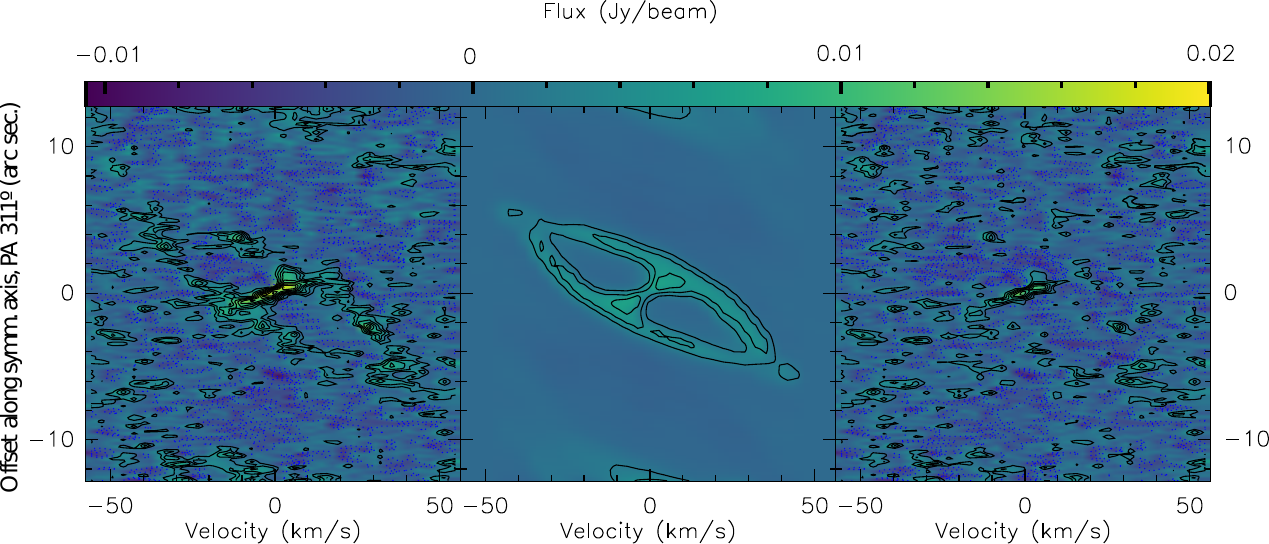}
      \caption{Position-velocity diagrams of observational data  from IRAM-NOEMA (left), the model (centre), and residuals (observation - model; right) from the \CdiecisieteO \, (top panels) and \CdieciochoO \, (bottom panels) \jdu \, maps.  Contours are drawn at 5\,mJy\,beam$^{-1}$ and 2\,mJy\,beam$^{-1}$ intervals respectively.}
         \label{c17&18omap}
 \end{figure*}

Thanks to the self-similar (i.e. Hubble-like) expansion of the nebula, where the velocity variable is proportional to radial position, we can use these P-V diagrams to show different cuts of the nebular structure. In Figure \ref{13comap} we use this property to compare the observational data and the simulated map. A clear asymmetry in the central area shows that our model lacks a bright spot in the nebula's equator, slightly displaced towards positive velocities. Although the physical properties of the central cylinder in our model attempt to reproduce the behaviour of this spot across all lines, it is clearly not possible with the limited resolution we have of this area. As a consequence, our model produces slightly higher peaks in lower lines and too narrow peaks in higher-excitation lines. A similar pattern is found in the \CdiecisieteO \, and \CdieciochoO \, maps as seen in Figure \ref{c17&18omap}.

While we can find other asymmetries in the observational maps, most of them are also reproduced by the model after going through the `interferometrisation', including thickness variations between analogue points of the shell despite its perfect symmetry, or the instrumental emission and absorption artefacts. Also, none of them are as strong as this central spot, which means this is, without any doubt, a real feature in the nebula with different physical variables than those given to the central cylinder.

As a result, the intensity of the central area becomes the main source of disagreement between both datasets, also in the line profiles. Concerning the \doceCO \, line profiles, we see how the central peak of the model is overestimated in the lower transitions up to \jcc. For higher energy transitions, however, the situation is the opposite. Even though this difference in intensity in the very centre is within uncertainties given the noise margin, it is important to notice that the peak reproduced by our model is much narrower than that observed. This shows that an additional unmodelled structure must be present in the nebula covering a wider range of central velocities, either due to its size or to the turbulence of its gas, and that it is slightly redshifted.

In all, our model simulates very accurately the observational data of CO isotopologues. Line profiles and interferometric maps helped build a robust model of the structures that dominate the emission in these lines.  However, we can improve the determination of the characteristics of the nebula by also modelling the emission from other molecular species.

\subsection{\hcop}

As less dense areas with higher temperatures are invisible in the lower transitions of CO maps, a proper description of the blobs' velocity law and position is impossible without the help of \hcop\ and HCN maps.
The best adjustment for these turbulent areas, while keeping a relatively simple description, is a constant value of 18\,\kms\ for the turbulence, but centred on different radial velocities. The 0\seca3 of the sphere farther away from the centre of the nebula are set at 15\,\kms\ of radial expansion velocity, while the closer part is set at 55\,\kms. The accuracy of this front-like velocity distribution, as seen by its coverage on the synthetic map, implies the presence of an ongoing shock on the fore parts between the ejected gas and previously existing material.

\begin{figure*}
    \includegraphics[width=\textwidth]{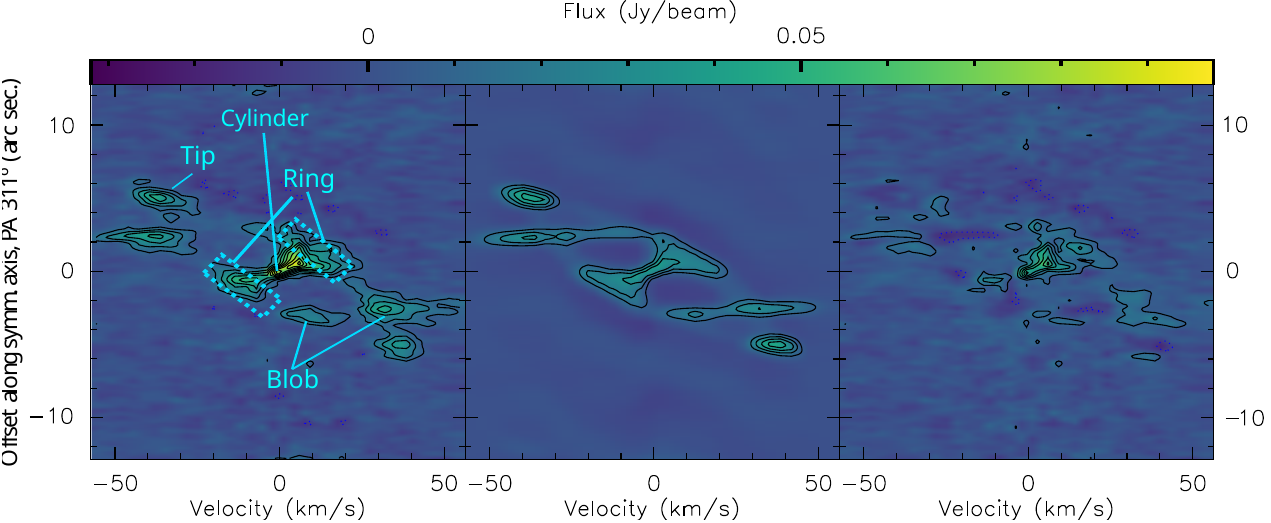}
      \caption{Position-velocity diagrams of observational data from IRAM-NOEMA (left), the model (centre), and residue (observation-model; right) from the \hcop \, \jdu \, map. Contours are drawn at 10\,mJy\,beam$^{-1}$ intervals. The structures that dominate the emission are annotated in the observational data panel.}
         \label{hcopmap}
 \end{figure*}

Besides the blobs, we can see from the maps that \hcop \, emission is bright on the tips and central area of the nebula, but the shell also seems to have a relevant \hcop\ abundance, particularly intense at its equator. This distribution prompted the addition of the ring structure, which only has a role different from that of the shell in the abundance of this molecule. As seen in Table \ref{tab:Abundances}, to properly reproduce the observations, we needed to introduce
different relative abundances for the different structures, contrary to CO species, where the same abundance is used across the model. The synthetic map resulting from our model, as well as the comparison to the observational data, can be found in Figure \ref{hcopmap}.

\begin{figure*}
    \includegraphics[width=\textwidth]{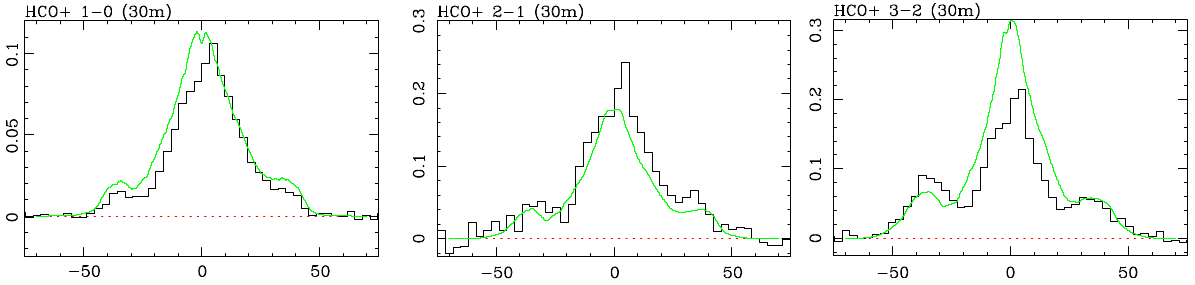}
      \caption{Comparison of line profiles obtained from IRAM-30M observations (black) and model reproduction (green) fot \hcop \, lines in \tmb\ (K) vs LSR velocity (\kms).}
         \label{hcoplines}
 \end{figure*}

Once again, in the data we find a bright spot slightly displaced towards positive velocities in the central part of the nebula not reproduced by our model, with a clear effect also on the line profiles as seen in Figure \ref{hcoplines}. The attempt at compensating for this emission makes the adjustment of the central peak less precise than those of the side peaks, just as it happens with \doceCO.

\begin{figure*}
    \includegraphics[width=\textwidth]{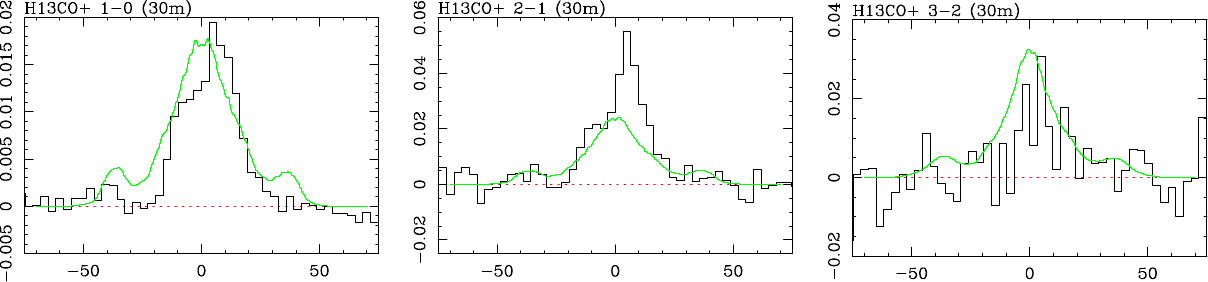}
      \caption{Comparison of line profiles obtained from IRAM-30M observations (black) and model reproduction (green) for \HtreceCOp \, lines in \tmb\ (K) vs LSR velocity (\kms).}
         \label{h13coplines}
 \end{figure*}

The model was also applied to \HtreceCOp\ for which the same abundance distribution across structures was deduced, just applying a constant factor with respect to the values of \hcop. The line profiles can be seen in Figure \ref{h13coplines}, where we note how the central peak is even more dominated by the bright spot than in \hcop\ and that it is displaced towards positive velocities.

Besides the slight inaccuracy of the central area of the nebula in most lines and maps, the overall modelling can be considered highly successful, as the physical variables remained the same from CO analysis, with only relative abundance needing to be determined for each species and structure to reproduce the map and all six line profiles. Therefore, the physical conditions used for the modelling are robust enough to be considered representative of those of the real nebula.

\subsection{HCN}

\begin{figure*}
    \includegraphics[width=\textwidth]{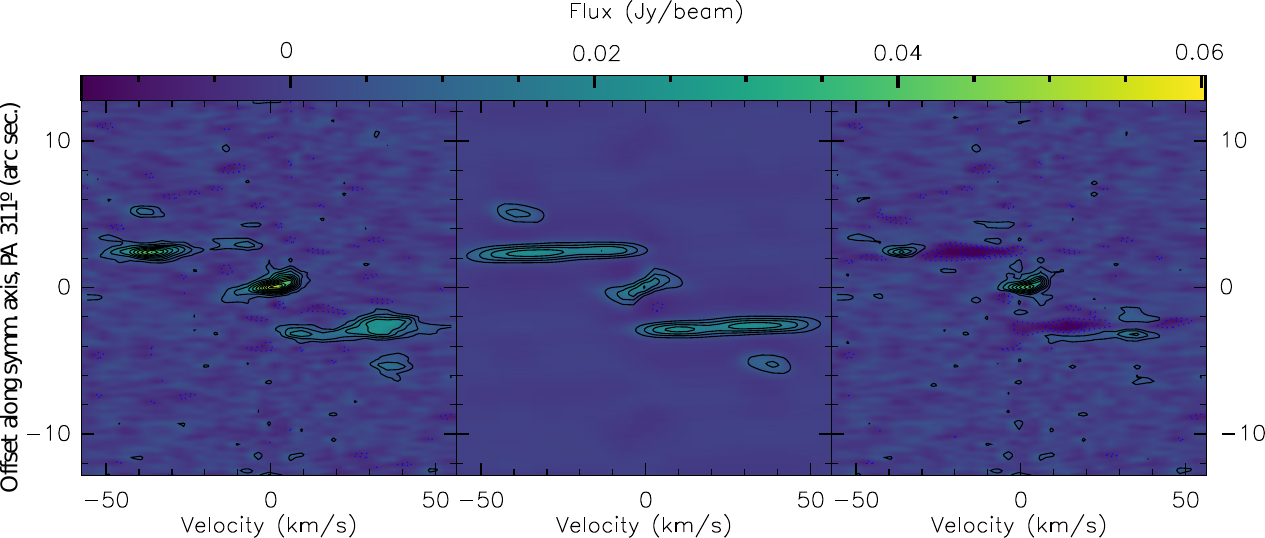}
      \caption{Position-velocity diagrams of observational data from IRAM-NOEMA (left), the model (centre), and residue (observation - model; right) from the HCN \jdu \, map. Contours are drawn at 5\,mJy\,beam$^{-1}$ intervals.}
         \label{hcnmap}
 \end{figure*}

The same process was followed for the HCN data. Compared to \hcop, the HCN is less extended across the nebula, with specific bright areas in the tips, blobs and central region, and less velocity dispersion overall. In HCN, the central bright region is located around the previously mentioned bright spot, and not found at all in the rest of the equator. As for \hcop, the relative abundance used in the modelling needs to be different in each structure to properly reproduce the observational data, being non-zero just for the inner structures. The interferometric \jdu\, map shows each of these three structures very clearly, as seen in Figure \ref{hcnmap}, where we show P-V diagrams for data, model, and data-model residuals.

\begin{figure*}
    \includegraphics[width=\textwidth]{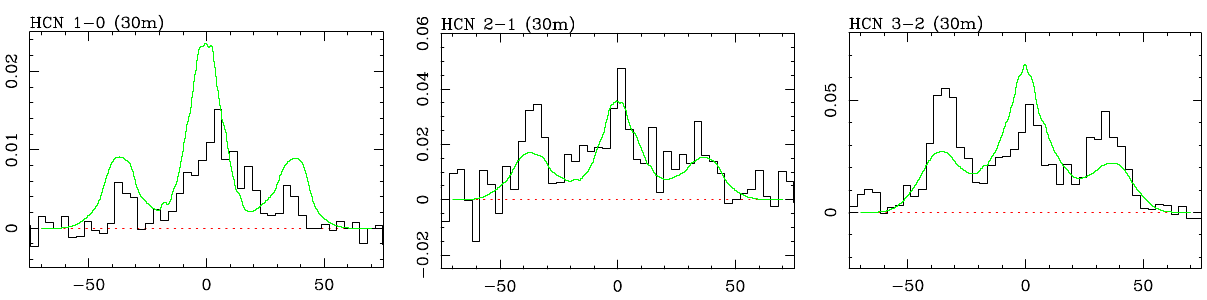}
      \caption{Comparison of line profiles obtained from IRAM-30m observations (black) and model reproduction (green) on HCN lines in \tmb\ (K) vs LSR velocity (\kms).}
         \label{hcnlines}
 \end{figure*}

However, the model predictions for the HCN line profiles do not reproduce the data as accurately as for other species. In particular, as seen in Figure \ref{hcnlines}, there are discrepancies in reproducing the side peaks across the three lines, an issue that was not found in any of the species modelled before. As this is an issue that only affects this species, and given the differences in the covered regions seen on maps, we expect this to be more likely due to the spatial distribution of HCN and the particular conditions linked to its emission, instead of related to wrongly adopted physical conditions for the overall nebula model.

\begin{figure*}
    \includegraphics[width=\textwidth]{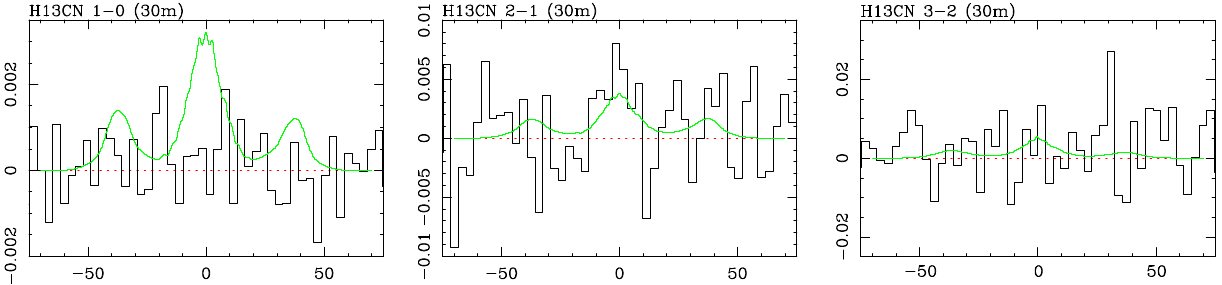}
      \caption{Comparison of line profiles obtained from IRAM-30m observations (black) and model reproduction (green) on \HtreceCN \, lines in \tmb\ (K) vs LSR velocity (\kms).}
         \label{h13cnlines}
 \end{figure*}

Once again, the model was applied to its \treceC\ isotopic substitution, \HtreceCN, but since the related observational data were dominated by noise, we have just opted for the same abundance ratio determined for \hcop.  In Figure \ref{h13cnlines}, we see that the model results are compatible with the data, considering the large data uncertainties. Only the \juc\ transition signal slightly exceeds the noise, but considering the overestimation in its $^{12}$C counterpart, the real emission would most likely be below the noise level.

\section{Discussion}
\subsection{Derived masses, kinetic energy, and scalar momentum}
From our 3D, radiative-transfer-enabled morpho-kinematic model, we derived the primary values of the nebula and their distribution across the object. After integrating all the model's cells, we obtained the following total values:
\begin{itemize}
    \item Total mass: 0.8$\pm$0.2\,\msun.
    \item Total kinetic energy: (6.5$\pm$1.3)$\times10^{45}$ erg.
    \item Total scalar linear momentum: (4.1$\pm$0.8)$\times10^{39}$ g·cm·s$^{-1}$.
\end{itemize}
We note that these values were calculated as an integration over all cells composing our model. Therefore, these results are not derived from any assumed abundance for a particular species but on the best-fit values of the physical parameters of the nebula computed over the 3D model simultaneously for every transition studied.
These values agree with previous studies \citep[][]{Bujarrabal_1998_Dynamics, Alcolea_2007, Li_2024, Khouri_2025} within a reasonable margin.
The highest discrepancy is found with this last study and the value they provide for the total mass of the nebula. \cite{Khouri_2025} find a total mass for \minky \, of 0.35\,\msun, assuming a distance of 2.6\,kpc, which would scale to 0.32\,\msun \, at our adopted distance. They compute this mass through an assumed temperature of 10\,K and a \CdieciochoO\, abundance with respect to hydrogen nuclei of 8.5$\times\ttt{-7}$. Although the assumed temperature is close to the one we obtain for the structures that dominate the \CdieciochoO\, emission, their assumed abundance differs quite a bit from ours, 6$\times\ttt{-7}$ with respect to the total number of collisional particles, the majority of them assumed to be molecular hydrogen. This means their assumed abundance is $\sim$2.8 times higher than ours, which is close to the ratio found between the total masses of both studies after taking into account the corrections for total gas mass. Given that \cite{Khouri_2025} is a survey-type study, where the same standard abundance and physical conditions are assumed for different objects, we consider our approach through detailed model fitting of observations more accurate.

A smaller but still significant difference in the total mass retrieved is found when comparing with the results from \cite{Li_2024}, where a total mass of 1.1\msun \, is reported. This study also presents the estimated mass per structure: 1.02\msun \, for the torus, which is the equivalent to our central cylinder and ring, 0.072\msun \, in the lobe walls (our shell) and 2.7\,\ttt{-4}\msun \, for each of the blobs, with no modelling of the tips. Even though the value for the total mass of the nebula is very compatible with our results, we see how the mass distribution is a strong source of disagreement when comparing these values to those presented in Table \ref{tab:dmstruct}. \cite{Li_2024} locates most of the mass in the central area, describing a high density torus and very narrow lobe walls. This distribution is valid for observations in visible wavelengths, but it is in striking contrast with radio observations, which show that these lobe walls must be considerably thicker. This is probably due to the (visible) scattered light being mostly sensitive to the internal side of these structures. We note that the total mass derived in the model by \cite{Li_2024} also depends on dust properties and the gas-to-dust ratio assumed (200 for all components of the nebula), which, given the mixed chemistry found in \minky\, might not be of standard characteristics. It is important to highlight that in all of these studies, the total mass estimated is a very significant portion of the initial mass of the star, resulting in no relevant changes in the conclusions derived for the death of its central star.

We can also visualise the distribution of these values across the nebula and its components. We provide the distribution of these three values by structure in table \ref{tab:dmstruct}, together with the time that radiation pressure would take to provide these linear momenta. As shown in some of the previously cited studies, the overall nebula presents a momentum excess that requires a mechanism more powerful than radiation pressure to be accounted for. We note that, as for the contribution of this momentum excess per structure, the outer and denser regions are the main contributors to this excess, in contrast with the warmer structures along the main axis, which do not show such an excess. However, given the strong collimation of these later ejections, we do not assume radiation pressure to be the process behind their formation. Nevertheless, this exercise shows that the launching mechanism that happened more recently, responsible for the axial components, should be significantly less powerful than the one that formed the bulk of the nebula. It is also worth noting that the mass contained by the spheres in our model is comparable to that estimated for the H$_2$ emission inside the lobes by \cite{Bujarrabal_1998_shocks}.

\begin{figure*}

    \includegraphics[width=\textwidth]{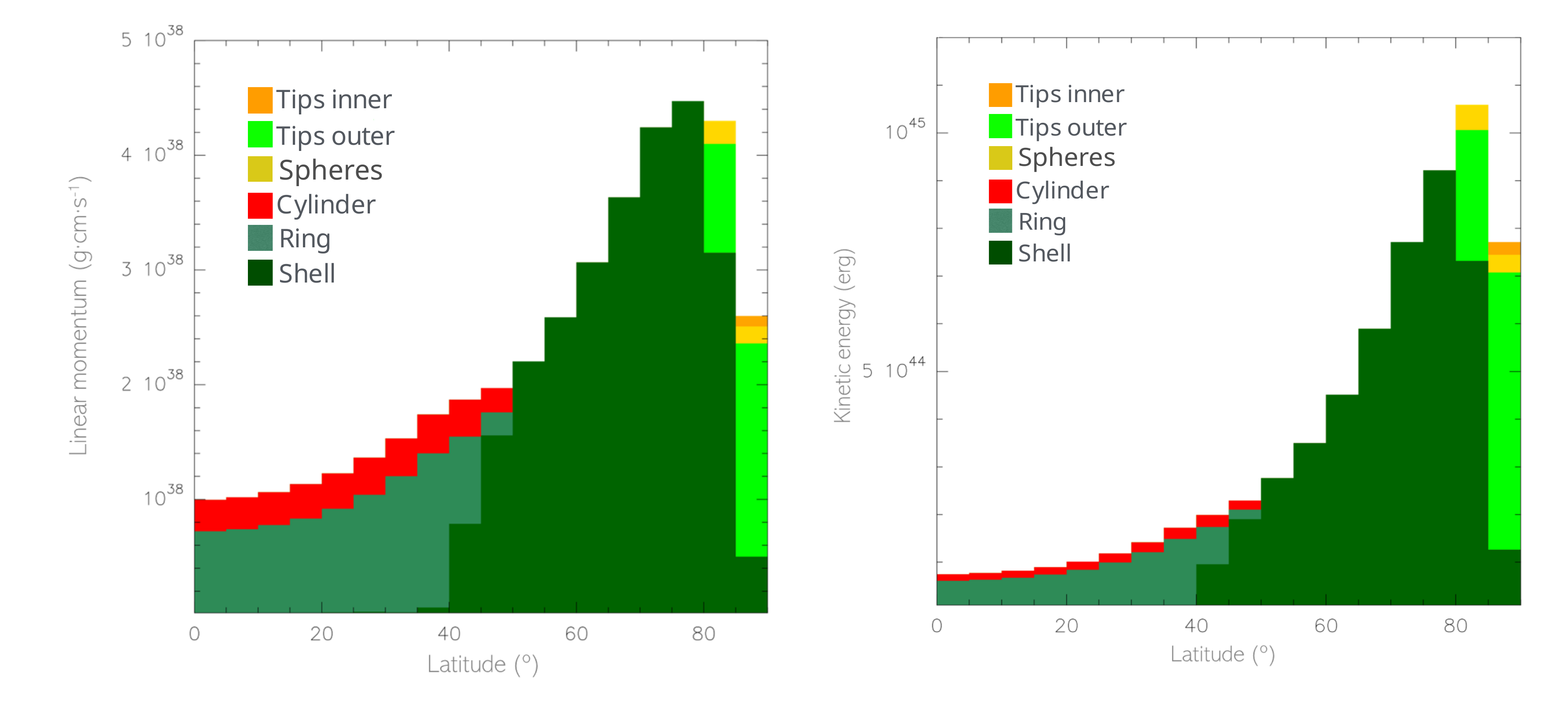}
      \caption{Distribution of scalar linear momentum (left) and kinetic energy (right) across the nebula as a function of latitude angle for the different structures.}
         \label{phi_distr_L}
 \end{figure*}

\begin{figure*}

     \includegraphics[width=170mm]{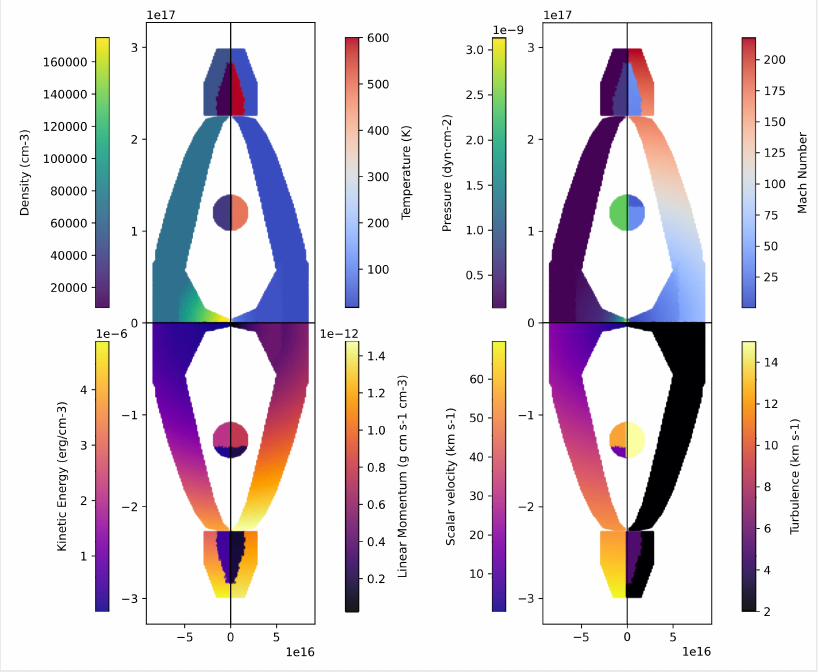}
      \caption{Distribution across the nebula of different physical values. In the left figure, the temperature (top right), density (top left), density of scalar linear momentum (bottom right), and density of kinetic energy (bottom left) are shown. In the right figure, the pressure (top left), Mach number (top right), scalar velocity (bottom left), and turbulence (bottom right) are shown. The axes are given in physical length units (centimetres). Note that 3.6$\times10^{16}$\,cm subtends 1\seca0 at the assumed distance of 2500\,pc.}
         \label{Plot_m192}
 \end{figure*}

We can also plot these values as a function of latitude. In Fig. \ref{phi_distr_L}, we can see the total amount of scalar linear momentum and kinetic energy with each structure's contribution in each latitude range. This figure clearly shows how material along the equator represents just a fraction of momentum and kinetic energy, whereas these quantities increase towards the polar direction and then drop at the very ends, where less material is detected. Finally, in Fig. \ref{Plot_m192}, we show the distribution of temperature, density, scalar linear momentum, kinetic energy, pressure, and Mach number across the plane containing the main axis of the nebula in our model. As axis symmetry and mirror symmetry with respect to the equatorial plane are kept in our model, a single quadrant is representative of the entire nebula.

\begin{table*}[]
        \caption{Derived values for each structure and the percentage they represent over the total value of the nebula.}
    \begin{tabular}{|c|c|c|c|c|}
    \hline
    Structure & Mass (\msun) | (\%) & Kinetic Energy (erg) | (\%)  & Scalar Lin. Mom. (g·cm·s$^{-1}$) | (\%) & Radiation press. timescale (a) \\ \hline
    Shell  & 0.4 | 52\% & 4.5\,\(\ttt{45}\) | 70\% & 2.6\,\ttt{39} | 64\% & 6.3\,\ttt{4}\\  \hline
    Ring   & 0.2 | 29\% & 8.0\,\(\ttt{44}\) | 12\% & 8.4\,\(\ttt{38}\) | 21\% & 2.0\,\ttt{4}  \\ \hline
    Cylinder & 0.1 | 17\% & 1.8\(\ttt{44}\) | 3\% & 3.0\,\(\ttt{38}\) | 7\% & 7.1\,\ttt{3}\\ \hline
    Tips out (each)  &  1.0 \(\ttt{-2}\) | 1\%   & 4.3\(\,\ttt{44}\) | 6.5\%   & 1.4\,\(\ttt{38}\) | 3\% & 3.4\,\ttt{3} \\ \hline
    Tips in (each)  &  4.0\,\(\ttt{-4}\) | 0.05\%   & 1.4\,\(\ttt{43}\) | 0.2\%   & 4.7\,\(\ttt{36}\) | 0.1\% & 1.1\,\ttt{2} \\ \hline
    Spheres (each)   &  1.9\,\(\ttt{-3}\) | 0.2\%   & 4.5\,\(\ttt{43}\) | 0.7\%   & 1.7\,\(\ttt{37}\) | 0.4\% & 4.1\,\ttt{2} \\ \hline
    \end{tabular}
    \tablefoot{The "radiation pressure timescale" is defined as the timescale needed for radiation pressure to provide the same linear momenta. All values present a $20\%$ error due to the abundance-density degeneracy.}
    \label{tab:dmstruct}
\end{table*}

We provide these total quantities and their distribution as a detailed constraint for HD/MHD models where this plotting is common \citep[see][]{Garcia-Segura_2005}. Unfortunately, there are no such models of \minky\, published currently, only the total values reported in \citealp{Bujarrabal_2001} are used when comparing with model results in \citealp{Garcia-Segura_2005} and \citealp{Blackman_2014}. As our results for these values are fairly compatible with the ones in \citealp{Bujarrabal_2001} we conclude that our work aligns with the results from these studies, where a common envelope ejection, or similar sudden event, is needed to reproduce these momentum and energy values. Therefore, we also avoid providing the reader with mass-loss rates. Since no precise ejection timescale is estimated for this source, we cannot present a value meaningful enough for its evolutionary process.

\subsection{Abundances}

The full final description of relative abundances per species and structure can be found in Table \ref{tab:Abundances}. The initial values of relative abundance with respect to collisional particles of the CO species assumed from the bibliography worked relatively well throughout the modelling process, thus requiring no significant changes. As for \hcop \, and HCN, we find a dichotomy between outer and inner structures. For outer ones, we also concluded values within the expected range in the typical CSEs of AGBs \citep[see][]{Pulliam_2011, Schoier_2013}. In contrast, the inner structures, especially the blobs and inner part of tips, present relative abundance values on the very upper limit or above. The strong presence of these molecules on the most dynamically active regions is yet another indicator of shocks \citep[][]{Viti_2002, Jorgensen_2004, James_2020}.

With this information on relative abundances with respect to collisional particles from our model, we derived the abundance ratios between the isotopologues of the studied species. We also estimated the error for the three main ratios as the variation needed for the difference to be significant in our model, while keeping the same physical variables.

\begin{itemize}
    \item \doceCO/\treceCO \,$= 30\pm$3;
    \item \CdiecisieteO/\CdieciochoO \,$= 1.6\pm$0.1.
    \item \HdoceCOp/\HtreceCOp \, $=10\pm$1.
\end{itemize}

Note that we can reasonably assume the same atomic isotopic ratio for \doceC/\treceC \, and $^{17}$O/$^{18}$O as those obtained through molecular species, since no chemical mechanism favouring certain isotopes (such as fractionation) is likely to be operating in \minky\,  (see section 5.3).The \CdiecisieteO/\CdieciochoO \, ratio matches the numbers obtained by line ratios once the frequency and Einstein coefficient corrections are applied to \CdiecisieteO \, and \CdieciochoO, as well as those estimated by previous studies \citep[][]{Alcolea_2022, Khouri_2025}.

\begin{figure*}[h]
    \includegraphics[width=\textwidth]{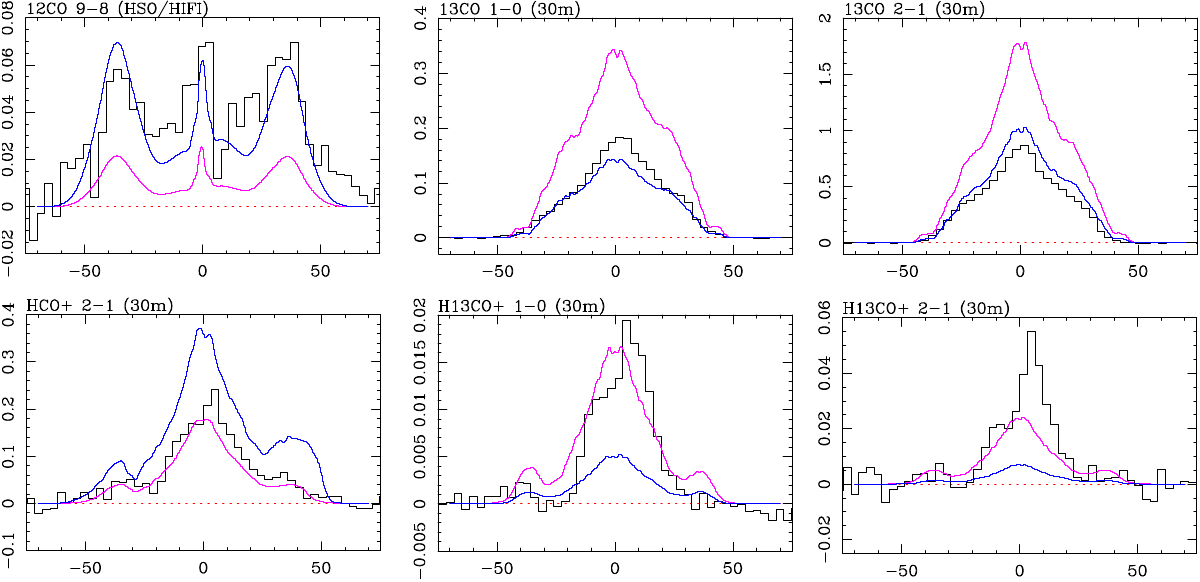}
      \caption{Comparison of model reproduction between \doceC/\treceC \, ratios of 30 (blue) and 10 (magenta). The vertical axes represent \tmb\ (K), while the horizontal axes are LSR velocity (\kms). Note how \doceCO\, and \treceCO\, lines (top panels) can only be successfully reproduced with a \doceC/\treceC\ ratio of 30, while \HdoceCOp\, and \HtreceCOp\, lines (bottom panels) require a value of ten.}
         \label{ratios}
 \end{figure*}

However, we find a strong discrepancy in the \doceC/\treceC \, ratio depending on the molecule used as a probe (CO or \hcop, see above). While both 30 and 10 are reasonable values for a mixed chemistry environment and not far from that previously estimated in the bibliography \citep[see][]{bujarrabal1997}, they are mutually incompatible in our model if applied to all the species. In Fig. \ref{ratios} we show the paradox of adopting either of these two $^{12}$C/$^{13}$C ratios in the adjustment of the different molecules by plotting some line profiles on the species used for probing this ratio: the optically thinnest line of \doceCO\, (\jno), both lines of \treceCO\ (\jdu\, and \juc), the best reproduced line of \HdoceCOp (\juc), and the less noisy lines of \HtreceCOp\, (\jdu\, and \juc). 
As can be seen in this figure, the adjustment of the CO lines fails miserably if we use the \doceC/\treceC\ ratio obtained from \hcop and vice versa. We note that the difference between both values (a factor of 3) is much larger than the uncertainty due to degeneracy between abundance and density.

A \doceC/\treceC\ ratio of ten was also applied to \HtreceCN, as the molecule traces an area very similar to \hcop. One can see that it is also compatible with the observational data, as seen in Fig. \ref{hcnlines}, where the model prediction would get hidden by the noise, with the exception of the \juc \, line already discussed in section 4.3.


\subsection{Physical implications}
Putting all of these results together, we can gain some insights into the physical circumstances that gave birth to this nebula. With the $^{17}$O/$^{18}$O abundance ratio of 1.6 obtained from CO isotopologues, we can derive an initial mass for the central post-AGB star. According to most nucleosynthesis models \citep[][]{De-Nutte_2017, Karakas_2014, Cristallo_2011, Stancliffe_2004}, this implies a stellar mass of around 1.7\msun \, at the beginning of the main sequence if we assume a solar metallicity, which despite not being measured, it is a reasonable assumption given the location of \minky\ in the galactic disc. This means that the nebular mass as obtained through our model would be around 70\% of the total mass loss expected throughout the entire AGB phase: for a 1.7\,\msun\ star we expect a white dwarf of 0.57$\pm$0.2\,\msun\ \citep[][]{Catalan_2008, Cummings_2018}. This initial mass should have also been enough for the central star to become C-rich at the end of the AGB phase, as it would have experienced enough third-dredge-up events to do so \citep[][]{Gronewegen_1995, Pardo_2007, Zhang_2013, Marigo_2020}. However, despite finding molecular species typical of C-rich environments, such as HCN, HNC or CS, the nebula shows a mixed chemistry with molecular species only found in O-rich environments, such as OH masers, SO, \sodos, and \water \, ices \citep[see][]{Alcolea_2022, seaquist1991, Eiroa_1983}. This result is compatible with previous studies such as \citealp{Khouri_2025}, where, despite the significant difference in \CdieciochoO\, abundance assumed in their work, the \CdiecisieteO/\CdieciochoO\, ratio, which is the value that nucleosynthesis models agree the most, is virtually the same.

The amount of mass ejected, 0.79\,\msun, relative to the initial mass, 1.7\,\msun, together with the expansion law, suggests that the formation of the nebula was triggered by a sudden event that also interrupted the AGB evolution and nucleosynthesis (and third-dredge-up events), as this would also explain its non-C-rich chemistry. Most likely, as proposed by previous studies \citep[][]{Alcolea_2007}, this event was a common envelope ejection, as these provide the energetics and a preferential axis for the ejection \citep[][]{Douchin_2015, Jones_2017}. We notice that this kind of pPNe, with a speculated sudden formation, is relatively common. Other examples are IRAS\,16342-3814, Hen 3--1475 or OH\,231.8+4.2 \citep[see][respectively.]{Sahai_2017, Khouri_2025, Alcolea_2001}

In addition, the warm, turbulent gas from the inner structures along the symmetry axis, tracing shocked gas, indicates the presence of later ejections, most likely related to polar jets, which are believed to indicate a binary system at its heart \citep[][]{Soker_2025}. If indeed we are witnessing the ejection of a common envelope, the fact that the total mass of the nebula amounts to the majority of the envelope of the AGB progenitor star would better align with the companion being a post-AGB star instead of a main sequence star (thus making the system a so-called double-degenerate) and therefore with the observed nebula being the product of the second common envelope ejection experienced by the system, as these seem to be substantially more massive than their single-degenerate counterparts \citep[][]{Santander-Garcia_2022}.

Since it seems clear that its physical properties differ quite significantly from the rest of the equatorial area, we can speculate about the nature of the central bright spot and whether it is also a fresh new ejection or has a different origin from all the other structures. However, we are limited to the spatial resolution of our maps, from which it is impossible to determine the actual position and size of this emission. Similar asymmetries have been found in other pPNe \citep[see][]{Castro-Carrizo_2002}, without an obvious explanation for their presence. Spatial resolution is certainly the most limiting aspect of our study, since the combined data is enough to derive the existence of a double-layered structure in the nebula, but we lack the information to reproduce a more realistic and accurate geometry in our model, with the failure in reproducing the HCN lines and the nature of the central spot as main consequences.

 The main challenge we face when it comes to putting these results in context comes from the values found for the $^{12}$C/$^{13}$C ratio. Although a ratio of 30 is in line with the interrupted AGB evolution via common envelope ejection hypothesis, as it is an intermediate ratio between those expected at the beginning and end of a thermally-pulsing AGB phase of a 1.7\,\msun\ star \citep[see F.R.U.I.T.Y. database\footnote{\href{https://fruity.oa-teramo.inaf.it/phys_modelli.pl}{https://fruity.oa-teramo.inaf.it/phys\_modelli.pl}} from][]{FRUITY_2016}, it is clearly different from the alternate value of 10 derived from \hcop \, and compatible with HCN. Both results seem to be very strong in our model, and the only way of reconciling them is by assuming different isotopic ratios for different structures and setting the \treceCO\ relative abundance 3 times higher only in the inner, warm structures. While this overestimates the \treceCO\ emission in the equatorial area, as our central cylinder is bigger than the central bright spot, it does not affect any other aspect of the \treceCO\ data we have. Therefore, we cannot conclude whether this is a molecular difference for the entire nebula or an isotopic ratio difference for carbon in certain structures, particularly affecting all the chemistry happening in them. In any case, we can ensure that this is not an opacity issue, as we base our results on the model properties, plus the smaller ratio would have been favoured for the most opaque lines, i.e. those of CO instead of \hcop. We can also discard isotopic fractionation as no part of our nebula has physical conditions with low enough temperature for it to happen, especially the ones traced by \hcop\ and HCN \citep[see][]{Roueff_2015}. In the same way, photodissociation is also not an option as the structure is delimited by the ballistic dynamics of the gas, with no sign of previously ejected material around it \citep[also discussed by][]{Li_2024}. Therefore, we theorise that structural difference, rather than molecular, is more likely to be the origin of these different $^{12}$C/$^{13}$C values. If we assume all the central structures, including the bright spot, come from more recent ejections, this would mean the central system kept evolving, including a decrease in the $^{12}$C/$^{13}$C isotopic ratio in the most recently ejected material. We do not dare to propose an exact mechanism for it, but very low ($\leq$ 5) $^{12}$C/$^{13}$C ratios are also found in other nebulae of explosion-like origin \citep[see][where possible mechanisms are discussed]{Kaminski_2017}. We note that a similar result was found in the pPN CRL 618 by \citep{Pardo_2007}, where a change in \doceC/\treceC\, ratio is reported, with a value of $\sim$40 in the halo decreasing to 15 in the core. This result is supported by other studies using different tracers for the isotopic ratio \citep[see][]{Wyrowski_2003, Wesson_2010, Lee_2013}. According to \cite{Lee_2013} this change occurred in a timespan of $\sim$450 years, a timescale very similar to that in \minky\, assuming the blobs as the main source of this disagreement in our study.

\subsection{Model limitations and sources of uncertainty}

\begin{figure*}[h]
    \includegraphics[width=\textwidth]{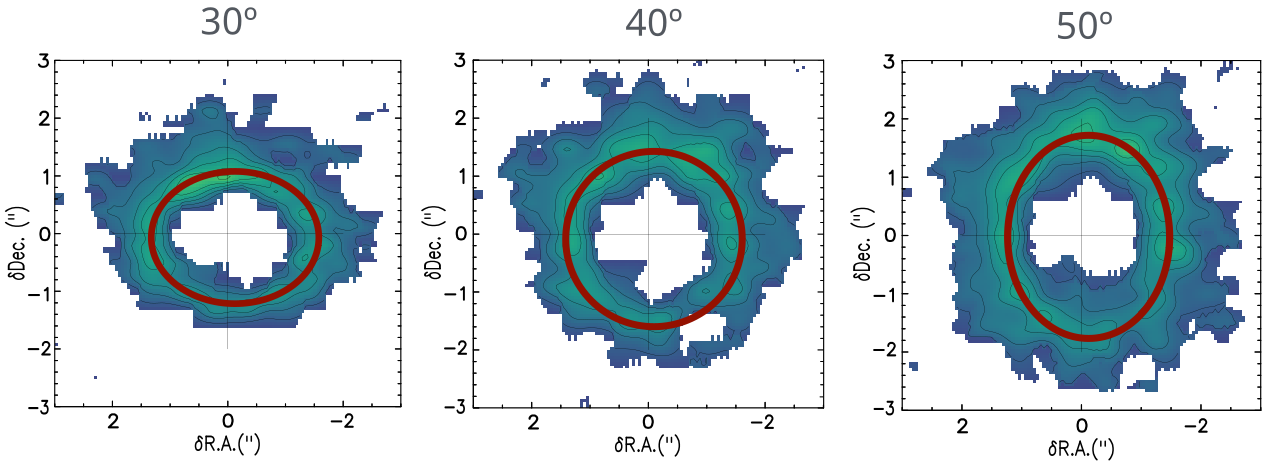}
      \caption{Cuts of the nebula performed perpendicularly to its symmetry axis taken at an intermediate position of the lobes for inclinations of 30º, 40º, and 50º with respect the plane of the sky once the projected velocities had been converted into positional points along the line of sight for the \treceCO\ \jdu \, map. In red is the fit around the data, showing the elongation or lack of.}
         \label{Angle}
 \end{figure*}

All of our best-fit physical values are supported by the strong alignment of the general characteristics of the nebula (total mass, kinetic energy, morphology, etc.) with previous studies \citep[][]{Solf_1994, Alcolea_2007, Alcolea_2022}. Among those, the inclination of the symmetry axis is the parameter that shows the largest discrepancy among studies, as discussed by \cite{Li_2024}. In this case, our inclination of 40º with respect to the plane of the sky (50º respect to the line of sight), results in the best-fit model in which revolution symmetry along the main axis is conserved, and a single radial gradient is enough for describing the velocity field. In particular, this angle is obtained by looking at the revolution symmetry of the lobes. Provided a velocity gradient to convert velocity information along the line of sight into positional information, cuts of the data at the correct angle will result in circular (revolution symmetry) images of the nebula. In Fig. \ref{Angle} we show the result of these cuts at different angles, showing elongations at other angles and a circular fit at the angle used for this modelling. We assume, at most, an error of $\pm$5 degrees, given the data noise, but further than that, revolution symmetry could not be assumed for the nebula. However, this is still an important source of uncertainty, as the real size will be affected. In the same way, the distance to the nebula, or the relative abundance of CO with respect to the colliding particles can change our results.

However, besides the uncertainties intrinsic to the observations, such as spatial resolution, axis inclination or distance to the nebula, the modelling method still imposes further limitations. The most important limitation is the very large number of variables in this modelling process and their interdependencies, which makes it virtually impossible for us to give exact error values for each of the estimated physical variables. However, this limitation is mitigated by the extended and varied amount of data used for this work, which heavily constrains our model. To check the extent of this robustness, an exploration of the parametric space has been carried for those pairs of physical variables that can produce the most degeneracy in the model reproduction. These pairs are Abundance-Density and Density-Temperature. In addition, each of these parameters has been explored individually while keeping the rest of the best-fit values fixed. These checks are performed on the \doceCO \, data, which are the ones used to define the best-fit values of the physical variables of the model, and to which the rest of abundances are referred. This exploration is also performed globally in our model, which means that the parameters are modified by the same percentage of their best-fit value in all structures at once, otherwise, the large number of combinations for each parameter and structure would make this checking process unfeasible. Due to the large number of parameters in the model, and their complex interdependences (e.g. mass and inclination with density and abundance), providing exact uncertainties for all values is far from trivial.

We find the density to have a $\pm10\%$ deviation margin with respect to its best-fit value before the model reproduction becomes incompatible with the observational results, including the observational data calibration errors discussed in section 2. However, this uncertainty becomes larger when taking into account the degeneracy with the abundance parameter, which can compensate for a change in density up to a 20\%. We note that this is not the only source of uncertainty for the density, where the volume of the structure and therefore the spatial resolution of maps also play a role. We estimate an uncertainty of 30\% of the ratio of the beam size to each structure size, which adds an extra 15\% for the largest structures and 30\% for the smallest to the density uncertainty. All of these errors are included in the uncertainty provided in Table \ref{tab:modelvar}.

For temperature the individual margin is of $\pm8\%$. We note that temperature is very constrained in the structures where \doceCO \, is optically thick, discarding a possible degeneracy of the temperature-density pairing. It is true that optically thin structures could present certain degeneracy; however, even the higher lines show a strong presence of optically thick structures (\jno\, line has a significant contribution of the central cylinder to the line profile), making the exploration of individual structures difficult to evaluate and overall inconsistent with the checking method used.

The $\pm20\%$ error on the abundance-density combination translates also into the \doceCO \, abundance and therefore to all other abundance values. This error, together with the individual errors accounting also for data calibration, are presented in Table \ref{tab:Abundances}. We highlight that this degeneracy on the abundance-density combination does not affect the abundance ratios since a change of density would require proportional changes in all other abundances. In the same way, the data calibration does not have any effect on the isotopic ratio values as observations (both single-dish and interferometric maps) of same molecule isotopologues where observed simultaneously. In the cases where the bandwidth was not enough to observe them at the same time, a third transition from another species common to both setups acts as calibrator. This way, the relative calibration is the same for those transitions, allowing the ratios to be as accurate as possible by minimising the relative error.

Regarding velocity, the general velocity law is also adopted from previous studies, and does not impose further uncertainty than that of the spectral resolution (3.25\,\kms). This does create a degeneracy in the possible velocity law of slower parts, i.e. central area of the nebula. The only structures whose velocity law has been determined throughout the modelling process are the spheres, where, due to their small size, the spatial resolution introduces an error in the determination of the shock-front limit of $\pm$0.\arcsec1, as inferred from the emission distribution in the different velocity channels of the maps.

All of these modelling errors and instrument contributions are also taken into account when presenting the error margins of the derived masses, kinetic energy and scalar momentum. However, not all sources of errors affect these final values. For instance, the error on the total density value derived from geometrical uncertainties cancels out with volume uncertainties when computing the total mass.

\section{Conclusions}
The new observational data presented in this paper provides the most complete and up to date view of \minky\, and consequently  motivated our detailed study and analysis of each structure of the nebula. We have presented a significant upgrade of the \shapemol \, code together with minor updates on the \SHAPE \, software as a whole. This upgrade provides a huge improvement to the capabilities of this modelling tool, expanding the previous range of physical conditions for \doceCO \, and \treceCO \, and adding ten new molecular species. The use of radiative transfer and LVG approximation provide an accurate reproduction of the observational data in many physical conditions, including optically thick cases. This sets \SHAPE \, to the standard of top radio telescopes and interferometers in the world, which are currently detecting and mapping not just CO but many other molecules of great impact in the study of astrophysical objects.

We brought this new update to the real case of \minky, making a complete morpho-kinematical study of the nebula. This is the most complete study of this object to date, collecting 23 line profiles and five interferometric maps to be reproduced simultaneously and giving strong constraints to the physical variables. As a result, we derived for the first time the existence of a two-layer structure along the preferential axis of the nebula together with its turbulent and shock-front features. In addition, we were able to study the chemical characteristics of the nebula, finding over-abundances of certain molecules and a discrepancy in the $^{12}$C/$^{13}$C isotopic ratio depending on the molecule used as a probe. Lastly, we present our results as a global distribution of the key physical variables across the nebula and as a function of latitude, resembling the products of HD/MHD models, hoping that in this way, comparison with these models will be more immediate.

All of these findings shed light on the origin of this nebula and, by extension, of the formation of other pPNe of this kind. Our analysis shows that the central star had its thermally pulsing AGB phase interrupted by an event that triggered the main nebular ejection and that the ejection happened at once. However, there are still many features that raise new questions about the formation and evolution of the nebula, including the apparent $^{12}$C/$^{13}$C ratio change during the post-AGB evolution and its link with the later collimated ejections as well as the nature of the central bright spot. We hope to answer these questions in future works, provided we obtain more spatially detailed observations.

\section{Data availability}

The electronic material produced in this study, i.e. \shapemol\ tables and \SHAPE\ model and script, can be found in \href{https://doi.org/10.5281/zenodo.19473007}{https://doi.org/10.5281/zenodo.19473007}

\begin{acknowledgements}

This work is part of the coordinated research project NEBULAe WEB, grants PID2019-105203GB-C21 and PID2019-105203GB-C22, funded by MICIU/AEI/10.13039/501100011033, and coordinated project MESON, grants PID2023-146056NB-C21 and PID2023-146056NB-C22, funded by MICIU/AEI/10.13039/501100011033 and ERDF/EU.
EM also thanks support from the IGN (MITMS, Spain) fellowship programme.

This paper has made use of data obtained with the IRAM 30m-MRT and NOEMA telescopes, and HSO/HIFI. The observations were carried out under project number W17BJ with the IRAM NOEMA Interferometer, projects number 050-15, 160-15 and 047-16 with the IRAM 30m telescope and projects GT1\_dteyssie\_1 and OT1\_vbujarra\_4 with the HSO/HIFI telescope. IRAM, the Institut de radioastronomie millimétrique, is an international research institute funded by the French Centre National de la Recherche Scientifique (CNRS), the German Max-Planck Gesellschaft (MPG) and the Spanish Instituto Geográfico Nacional (IGN). HSO, the Herschel Space Observatory, is an ESA space observatory with science instruments provided by European-led Principal Investigator consortia and with important participation from NASA. HIFI, the Heterodyne Instrument for the Far Infrared, has been designed and built by a consortium of institutes and university departments from across Europe, Canada, and the United States under the leadership of SRON (Netherlands Institute for Space Research), Groningen, The Netherlands, and with major contributions from Germany, France, and the US. Consortium members are: Canada: CSA, U.Waterloo; France: CESR, LAB, LERMA, IRAM; Germany: KOSMA, MPIfR, MPS; Ireland, NUI Maynooth; Italy: ASI, IFSI-INAF, Osservatorio Astrofisico di Arcetri-INAF; The Netherlands: SRON, TUD; Poland: CAMK, CBK; Spain: Observatorio Astronmico Nacional (IGN), Centro de Astrobiologa (CSIC-INTA). Sweden: Chalmers University of Technology-MC2, R SS \& GARD; Onsala Space Observatory; Swedish National Space Board, Stockholm University-Stockholm Observatory; Switzerland: ETH Zurich, FHNW; USA: Caltech, J.P.L., NHSC. This work has made use of GILDAS software for data reduction and analysis (\href{https://www.iram.fr/IRAMFR/GILDAS}{https://www.iram.fr/IRAMFR/GILDAS}). We also acknowledge the work by Ismael Domínguez Salamanca during his Master's Final Project.

\end{acknowledgements}

%
%

\bibliographystyle{aa}
\bibliography{export-bibtex.bib}

\onecolumn
\appendix

\section{Additional tabular material}

\begin{table*}[h]

        \caption{General characteristics of the tables provided with $k_v$ and $j_v$ coefficients calculated to allow \shapemol\ 's calculations.}
    \begin{tabular}{|c|c|c|c|c|}
    \hline

    Molecular & Temperature range (K)  & Density range (cm$^{-3}$)             & Relative abundance range                  & Number    \\ 
    species   & \& step (K) & \& step  & \& step                  & of tables \\ \hline
    \doceCO   & 5 --1000       & \ttt{2} -- \ttt{12}    & 2.3\,\ttt{-7} -- 1.2\,\ttt{-2} & 52 \\ 
              & 5              & $\sqrt[4]{10}$         & $\sqrt[NT-1]{5.3\,\ttt{4}}$  &    \\ \hline
    \treceCO   & 5 --1000       & \ttt{2} -- \ttt{12}    & 7.7\,\ttt{-9} -- 4.1\,\ttt{-4} & 52 \\ 
              & 5              & $\sqrt[4]{10}$         & $\sqrt[NT-1]{5.3\,\ttt{4}}$  &    \\ \hline
    \CdiecisieteO   & 5 --1000       & \ttt{2} -- \ttt{12}    & 4.6\,\ttt{-9} -- 2.4\,\ttt{-4} & 52 \\ 
              & 5              & $\sqrt[4]{10}$         & $\sqrt[NT-1]{5.3\,\ttt{4}}$  &    \\ \hline
    \CdieciochoO   & 5 --1000       & \ttt{2} -- \ttt{12}    & 2.3\,\ttt{-9} -- 1.2\,\ttt{-4} & 52 \\ 
              & 5              & $\sqrt[4]{10}$         & $\sqrt[NT-1]{5.3\,\ttt{4}}$  &    \\ \hline
    SiO   & 5 --1000       & \ttt{2} -- \ttt{12}    & 1.0\,\ttt{-12} -- 1.5\,\ttt{-7} & 56 \\ 
              & 5              & $\sqrt[8]{10}$         & $\sqrt[NT-1]{5.3\,\ttt{4}}$  &    \\ \hline
    CS   & 5 --1000       & \ttt{2} -- \ttt{12}    & 4.0\,\ttt{-12} -- 2.1\,\ttt{-7} & 52 \\ 
              & 5              & $\sqrt[8]{10}$         & $\sqrt[NT-1]{5.3\,\ttt{4}}$  &    \\ \hline
    \hcop   & 5 --1000       & \ttt{2} -- \ttt{12}    & 1.0\,\ttt{-10} -- 5.3\,\ttt{-6} & 52 \\ 
              & 5              & $\sqrt[16]{10}$         & $\sqrt[NT-1]{5.3\,\ttt{4}}$  &    \\ \hline              
    \HtreceCOp   & 5 --1000       & \ttt{2} -- \ttt{12}    & 3.3\,\ttt{-12} -- 1.8\,\ttt{-7} & 52 \\ 
              & 5              & $\sqrt[16]{10}$         & $\sqrt[NT-1]{5.3\,\ttt{4}}$  &    \\ \hline              
    HCN   & 5 --1000       & \ttt{2} -- \ttt{12}    & 1.0\,\ttt{-12} -- 1.0\,\ttt{-5} & 76 \\ 
              & 5              & $\sqrt[16]{10}$         & $\sqrt[NT-1]{\ttt{7}}$  &    \\ \hline
    \HtreceCN   & 5 --1000       & \ttt{2} -- \ttt{12}    & 3.3\,\ttt{-14} -- 3.3\,\ttt{-7} & 76 \\ 
              & 5              & $\sqrt[16]{10}$         & $\sqrt[NT-1]{\ttt{7}}$  &    \\ \hline
    HNC   & 5 --1000       & \ttt{2} -- \ttt{12}    & 3.3\,\ttt{-13} -- 3.3\,\ttt{-6} & 76 \\ 
              & 5              & $\sqrt[16]{10}$         & $\sqrt[NT-1]{\ttt{7}}$  &    \\ \hline
    \nnhp   & 5 --1000       & \ttt{2} -- \ttt{12}    & 1.0\,\ttt{-13} -- 1.0\,\ttt{-6} & 76 \\ 
              & 5              & $\sqrt[16]{10}$         & $\sqrt[NT-1]{\ttt{7}}$  &    \\ \hline              
    \end{tabular}
    \label{tab:shapemol1}
\end{table*}

\begin{table}[h]

        \caption{Line profile integrated areas and noise errors, in Jy\,\kms, for the molecular species / transitions observed with the IRAM 30\,m MRT and NOEMA interferometer. See Section 2 for discussion on observational data calibration.}
    \begin{tabular}{|l|l|r|r|}
    \hline

    Species & Transition & 30\,m MRT & NOEMA \\ \hline
    \doceCO        & \juc & 141.8$\pm$0.3 & \\
                   & \jdu & 422.7$\pm$0.3 & \\
    \treceCO       & \juc &  38.2$\pm$0.1 & \\
                   & \jdu & 171.1$\pm$0.3 & 174.0$\pm$0.1 \\
    \CdiecisieteO  & \juc &   2.3$\pm$0.1 & \\
                   & \jdu &  15.7$\pm$0.4 & 10.1$\pm$0.1 \\
    \CdieciochoO   & \juc &   1.4$\pm$0.1 & \\
                   & \jdu &   8.7$\pm$0.3 & 6.6$\pm$0.1 \\
    \hcop          & \juc &  17.6$\pm$0.1 & \\
                   & \jdu &  40.3$\pm$0.3 & 51.8$\pm$0.1 \\
                   & \jtd &  41.2$\pm$0.5 & \\
    \HtreceCOp     & \juc &   2.6$\pm$0.1 & \\
                   & \jdu &   5.7$\pm$0.3 & \\
                   & \jtd &   3.0$\pm$0.6 & \\
    HCN            & \juc &   2.1$\pm$0.1 & \\
                   & \jdu &   8.6$\pm$0.6 &   7.7$\pm$0.1\\
                   & \jtd &  13.7$\pm$0.5& \\
    \HtreceCN      & \juc & $<$0.3 & \\
                   & \jdu & $<$1.2 & \\
                   & \jtd & $<$2.1 & \\      \hline
    \end{tabular}
    \label{tab:linefluxes}
\end{table}

\section{Interferometric maps}

\begin{figure*}[h]

    \includegraphics[width=\textwidth]{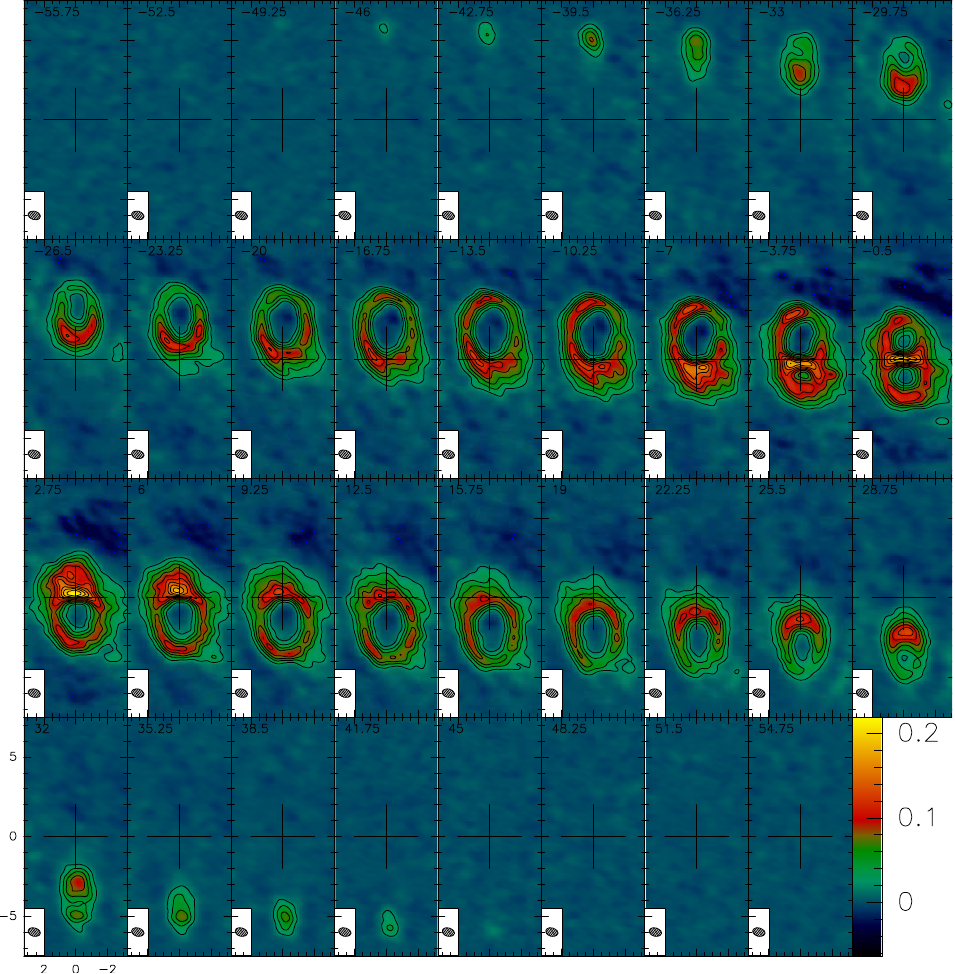}
      \caption{Interferometric maps obtained with IRAM NOEMA for the \treceCO\ \jdu\ transition for the different 3.25\,\kms -wide velocity channels. The central LSRK velocity (in \kms) for each channel is shown in the upper left corner of the corresponding panel. Note that these maps have been rotated 49º clockwise so the symmetry axis of the nebula is in the vertical direction. The Field of view is 6\seca5$\times$15\seca0. Clean-beam size is 0\seca51$\times$0\seca78, with a PA of 76º, and is displayed in the inset at the lower-left corner. Contours are drawn at 20 mJy\,beam$^{-1}$ intervals. The false-colour intensity scale displayed in Jy\,beam$^{-1}$ is also indicated right of the last panel.}
         \label{ratios1}
 \end{figure*}

\begin{figure*}[h]

    \includegraphics[width=\textwidth]{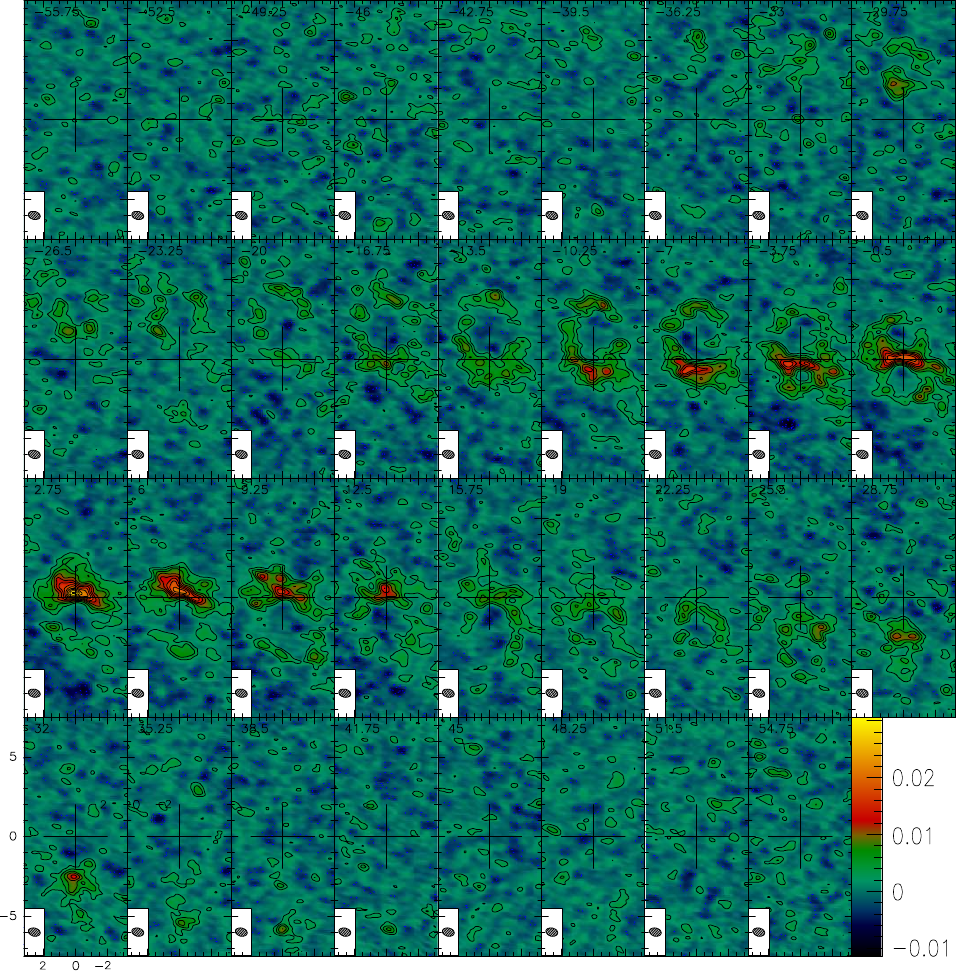}
      \caption{Same as Fig.\,\ref{ratios1} for the \CdiecisieteO\ \jdu\ transition and 2.5 mJy\,beam$^{-1}$ intervals for contours. Clean-beam size is 0\seca51$\times$0\seca76, with a PA of 76º.}
         \label{ratios2}
 \end{figure*}

 \begin{figure*}[h]

     \includegraphics[width=\textwidth]{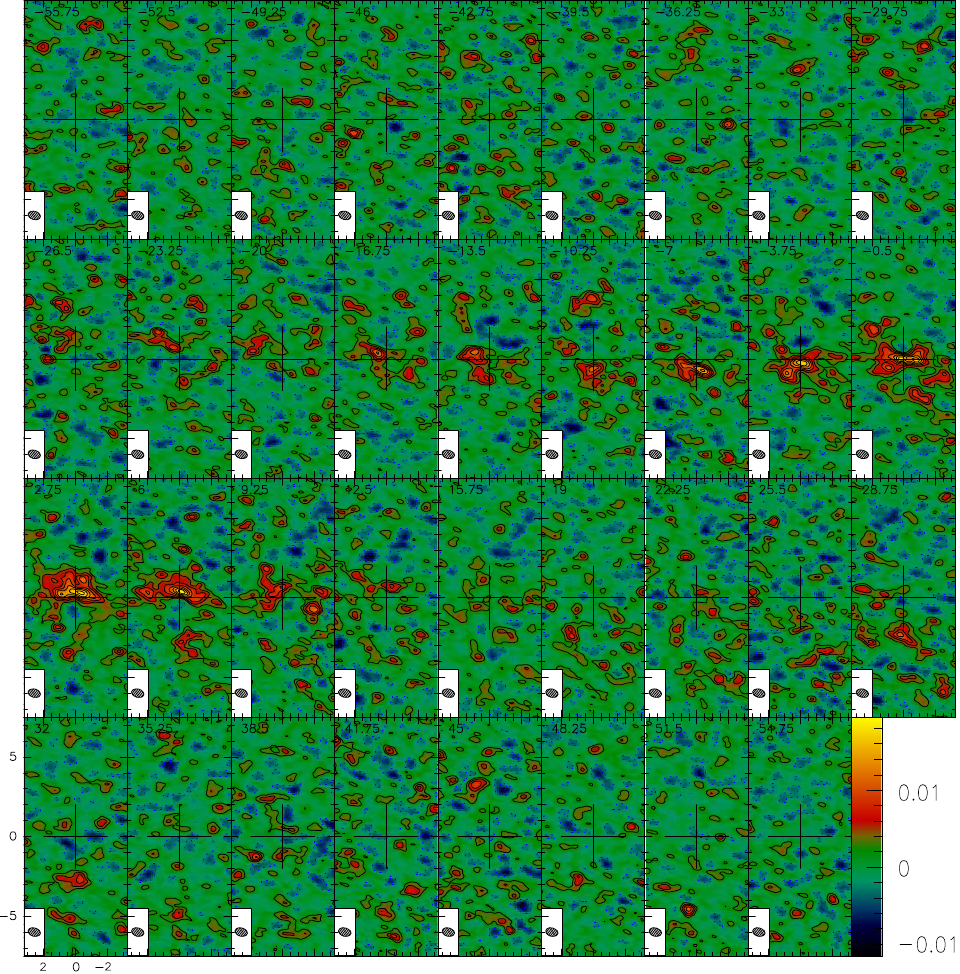}
      \caption{Same as Fig.\,\ref{ratios1} for the \CdieciochoO\ \jdu\ transition and 2.5 mJy\,beam$^{-1}$ intervals for contours. Clean-beam size is 0\seca51$\times$0\seca78, with a PA of 76º.}
         \label{ratios3}
 \end{figure*}

 \begin{figure*}[h]

     \includegraphics[width=\textwidth]{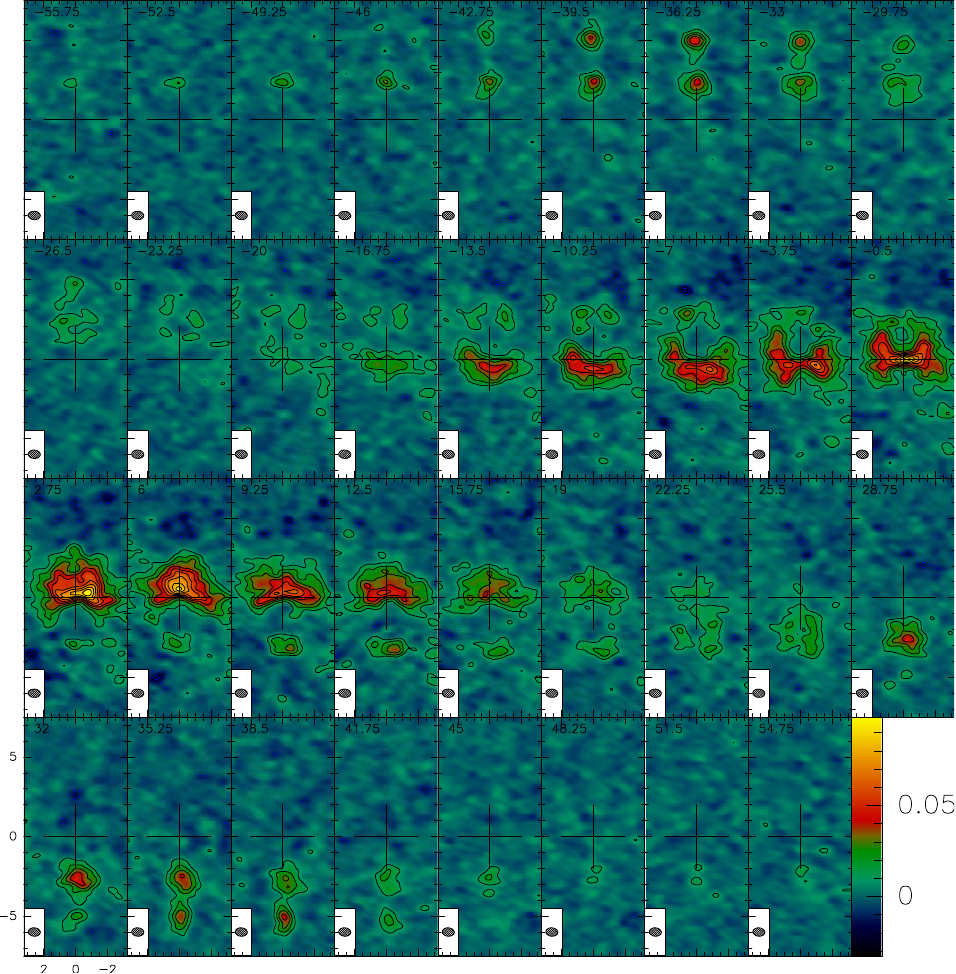}
      \caption{Same as Fig.\,\ref{ratios1} for the \hcop\ \jdu\ transition and 10 mJy\,beam$^{-1}$ intervals for contours. Clean-beam size is 0\seca53$\times$0\seca76, with a PA of 86º.}
         \label{ratios4}
 \end{figure*}

 \begin{figure*}[h]

     \includegraphics[width=\textwidth]{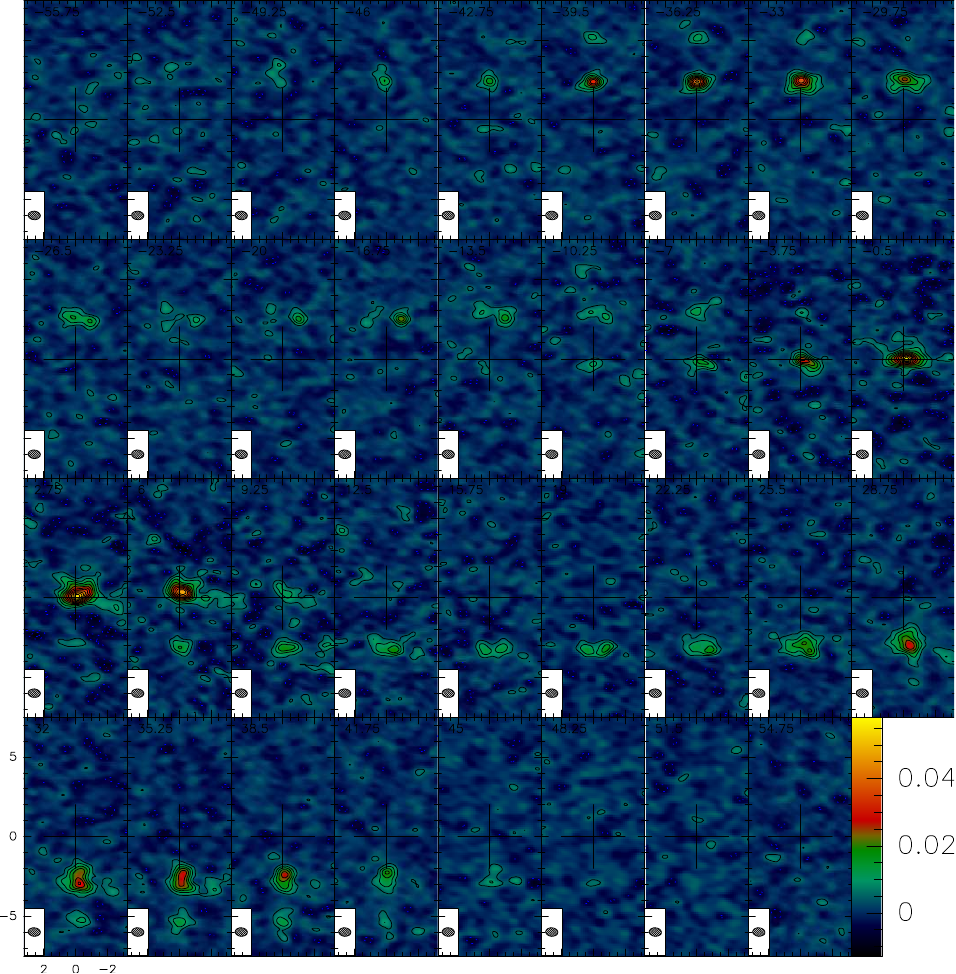}
      \caption{Same as Fig.\,\ref{ratios1} for the HCN \jdu\ transition and 5 mJy\,beam$^{-1}$ intervals for contours. Clean-beam size is 0\seca53$\times$0\seca77, with a PA of 86º.}
         \label{ratios5}
 \end{figure*}

\begin{figure*}[h]
    \includegraphics[width=\textwidth]{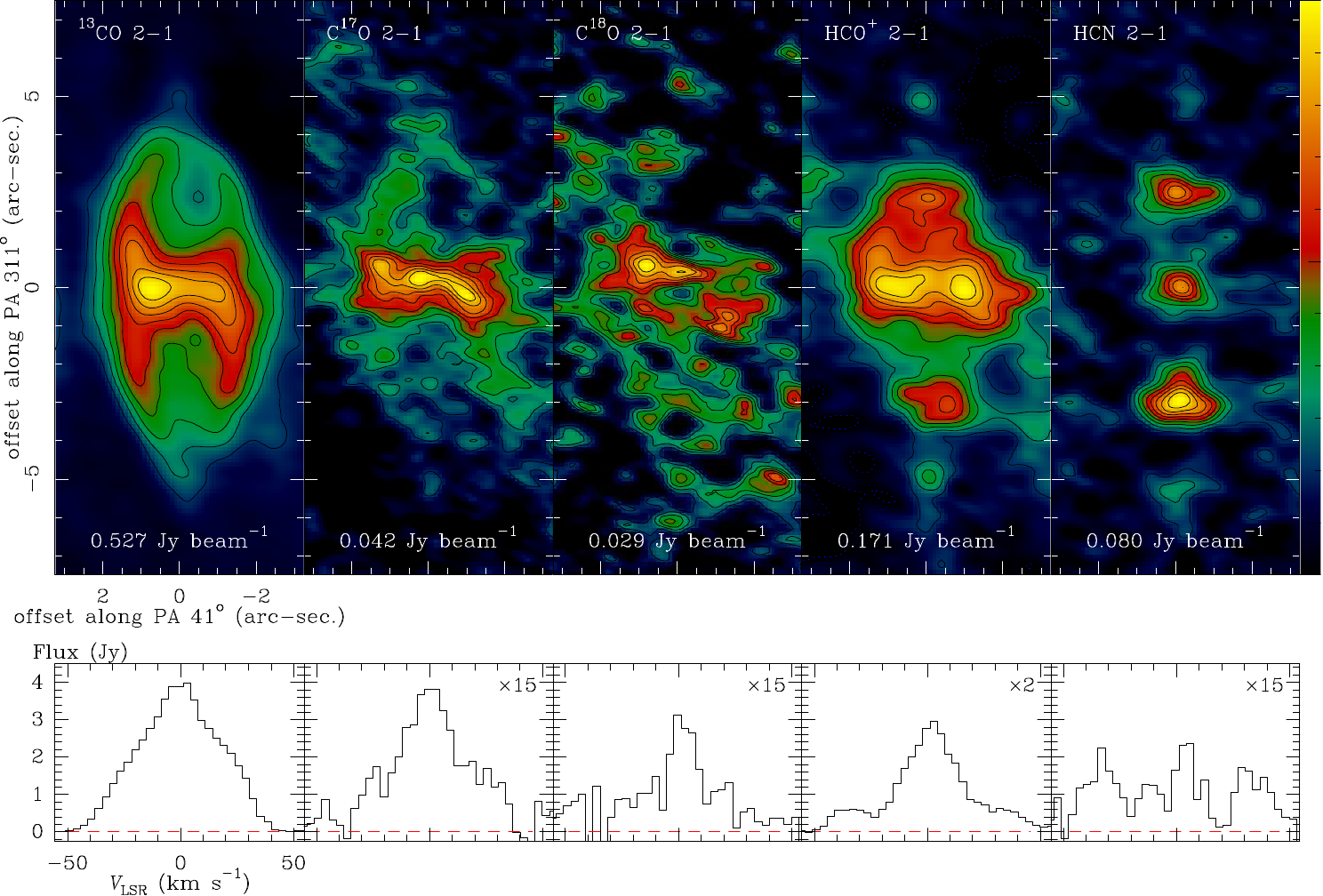}
      \caption{Moment zero maps (top panels) and integrated emission profiles (bottom panels) for the lines observed with the NOEMA interferometer: from left to right, \treceCO\ \jdu, \CdiecisieteO\ \jdu, \CdieciochoO\ \jdu, \hcop\ \jdu, and HCN \jdu. For the moment zero maps, the intensity is given in Jy\,beam$^{-1}$. Contours are drawn at 10\% to 90\% by intervals of 10\% of the maximum in the respective map. The contour step, 10\% of maximum, in Jy\,beam$^{-1}$, is given at the bottom of the panel. The colour scale is fixed between --10\% and 100\% of the maximum emission. Offsets are given along the major symmetry axis of the nebula (the maps have been rotated by 49\arcdeg\ so the revolution axis of the nebula, at PA 311\arcdeg\ is in the vertical direction). Integrated emission profile fluxes are given in Jy vs. {\em LSR} velocities in \kms. To keep the same flux scale in all five panels, spectra other than that of \treceCO\ \jdu\ have been multiplied by the factor indicated in the top right corner.}
         \label{mom0x5}
 \end{figure*}

\end{document}